# Physical Principles for Scalable Neural Recording

◁Adam H. Marblestone,[1,2] ◁Bradley M. Zamft,[3] Yael G. Maguire,[3,4] Mikhail G. Shapiro,[5] Thaddeus R. Cybulski,[6] Joshua I. Glaser,[6] Dario Amodei,[7] P. Benjamin Stranges,[3] Reza Kalhor,[3] David A. Dalrymple,[1,8,9] Dongjin Seo,[10] Elad Alon,[10] Michel M. Maharbiz,[10] Jose M. Carmena,[10,11] Jan M. Rabaey,[10] Edward S. Boyden,▷[9,12] George M. Church,▷[1,2,3] and Konrad P. Kording ▷[13,14]

◁ Joint first authors   ▷ Joint last authors   **1** Biophysics Program, Harvard Univ., Boston, MA 02115, USA   **2** Wyss Institute for Biologically Inspired Engineering at Harvard Univ., Boston, MA 02115, USA   **3** Dept. of Genetics, Harvard Medical School, Boston, MA 02115, USA   **4** Plum Labs LLC, Cambridge, MA 02142, USA   **5** Division of Chemistry and Chemical Engineering, California Institute of Technology, Pasadena, CA 91125, USA   **6** Interdepartmental Neuroscience Program, Northwestern Univ., Chicago, IL 60611, USA   **7** Department of Radiology, Stanford University, Palo Alto, CA 94305, USA   **8** Nemaload, San Francisco, CA 94107, USA   **9** Media Laboratory, Massachusetts Institute of Technology, Cambridge, MA 02139, USA   **10** Dept. of Electrical Engineering and Computer Sciences, Univ. of California at Berkeley, Berkeley, CA 94720, USA   **11** Helen Wills Neuroscience Institute, Univ. of California at Berkeley, Berkeley, CA 94720, USA   **12** Depts. of Brain and Cognitive Sciences & of Biological Engineering, Massachusetts Institute of Technology, Cambridge, MA 02139, USA   **13** Depts. of Physical Medicine and Rehabilitation & of Physiology, Northwestern Univ. Feinberg School of Medicine, Chicago, IL 60611, USA   **14** Sensory Motor Performance Program, The Rehabilitation Institute of Chicago, Chicago, IL 60611, USA

Correspondence to: adam.h.marblestone (at) gmail.com

> " To understand in depth what is going on in a brain, we need tools that can fit inside or between neurons and transmit reports of neural events to receivers outside. We need observing instruments that are local, nondestructive and noninvasive, with rapid response, high band-width and high spatial resolution… There is no law of physics that declares such an observational tool to be impossible. "
>
> Freeman Dyson, *Imagined Worlds*, 1997

## Abstract

Simultaneously measuring the activities of all neurons in a mammalian brain at millisecond resolution is a challenge beyond the limits of existing techniques in neuroscience. Entirely new approaches may be required, motivating an analysis of the fundamental physical constraints on the problem. We outline the physical principles governing brain activity mapping using optical, electrical, magnetic resonance, and molecular modalities of neural recording. Focusing on the mouse brain, we analyze the scalability of each method, concentrating on the limitations imposed by spatiotemporal resolution, energy dissipation, and volume displacement. Based on this analysis, all existing approaches require orders of magnitude improvement in key parameters. Electrical recording is limited by the low multiplexing capacity of electrodes and their lack of intrinsic spatial resolution, optical methods are constrained by the scattering of visible light in brain tissue, magnetic resonance is hindered by the diffusion and relaxation timescales of water protons, and the implementation of molecular recording is complicated by the stochastic kinetics of enzymes. Understanding the physical limits of brain activity mapping may provide insight into opportunities for novel solutions. For example, unconventional methods for delivering electrodes may enable unprecedented numbers of recording sites, embedded optical devices could allow optical detectors to be placed within a few scattering lengths of the measured neurons, and new classes of molecularly engineered sensors might obviate cumbersome hardware architectures. We also study the physics of powering and communicating with microscale devices embedded in brain tissue and find that, while radio-frequency electromagnetic data transmission suffers from a severe power–bandwidth tradeoff, communication via infrared light or ultrasound may allow high data rates due to the possibility of spatial multiplexing. The use of embedded local recording and wireless data transmission would only be viable, however, given major improvements to the power efficiency of microelectronic devices.

## 1 INTRODUCTION

Neuroscience depends on monitoring the electrical activities of neurons within functioning brains [1–3] and has advanced through steady improvements in the underlying observational tools. The number of neurons simultaneously recorded using wired electrodes, for example, has doubled every seven years since the 1950s, currently allowing electrical observation of hundreds of neurons at sub-millisecond timescales [4]. Recording techniques have also diversified: activity-dependent optical signals from neurons endowed with fluorescent indicators can be measured by photodetectors, and radio-frequency emissions from excited nuclear spins allow the construction of magnetic resonance images modulated by activity-dependent contrast mechanisms. Ideas for alternative methods have been proposed, including the direct recording of neural activities into information-bearing biopolymers [5–7].

Each modality of neural recording has characteristic advantages and disadvantages. Multi-electrode arrays enable the recording of ∼250 neurons at sub-millisecond temporal resolutions. Optical microscopy can currently record ∼100 000 neurons at a 1.25 s timescale in behaving larval zebrafish using light-sheet illumination [8], or hundreds to thousands of neurons at a ∼100 ms timescale in behaving mice using a 1-photon fiber scope [9]. Magnetic resonance imaging (MRI) allows non-invasive whole brain recordings at a 1 s timescale, but is far from single neuron spatial resolution, in part due to the use of hemodynamic contrast. Finally, molecular recording devices have been proposed for scalable physiological signal recording but have not yet been demonstrated in neurons [5–7].

Figure 1 illustrates the recording modalities studied here. While further development of these methods promises to be a crucial driver for future neuroscience research [10], their fundamental scaling limits are not immediately obvious. Furthermore, inventing new technologies for scalable neural recording requires a quantitative understanding of the engineering problems that such technologies must solve, a landscape of constraints which should inform design decisions.

**Figure 1.** Four generalized neural recording modalities. (a) *Extracellular electrical recording* probes the voltage due to nearby neurons. (b) *Optical microscopy* detects light emission from activity-dependent indicators. In two-photon laser scanning microscopy, shown here, an excitation beam at 2× the peak excitation wavelength of the fluorescent indicator is scanned across the sample, while an integrating detector captures the emitted fluorescence. (c) *Magnetic resonance imaging* detects radio-frequency magnetic induction signals from aqueous protons, after weak thermal alignment of the proton spins by a static magnetic field. A resonant radio-frequency pulse tips the spins into a plane perpendicular to the static field, causing the net magnetization to precess. The resulting signals are affected by the local chemical and magnetic environment, which can be altered dynamically by imaging agents in response to neural activity. Activity-dependent contrast agents are necessary to transduce neural activity into an MRI readout, whereas current functional MRI methods rely on blood oxygenation signals which cannot reach single-neuron resolution. (d) *Molecular recording* devices have been proposed, in which a "ticker tape" record of neural activity is encoded in the monomer sequence of a biomolecular polymer – a form of nano-scale local data storage. This could be achieved by coupling correlates of neural activity to the nucleotide misincorporation probabilities of a DNA or RNA polymerase as it replicates or transcribes a known DNA strand.

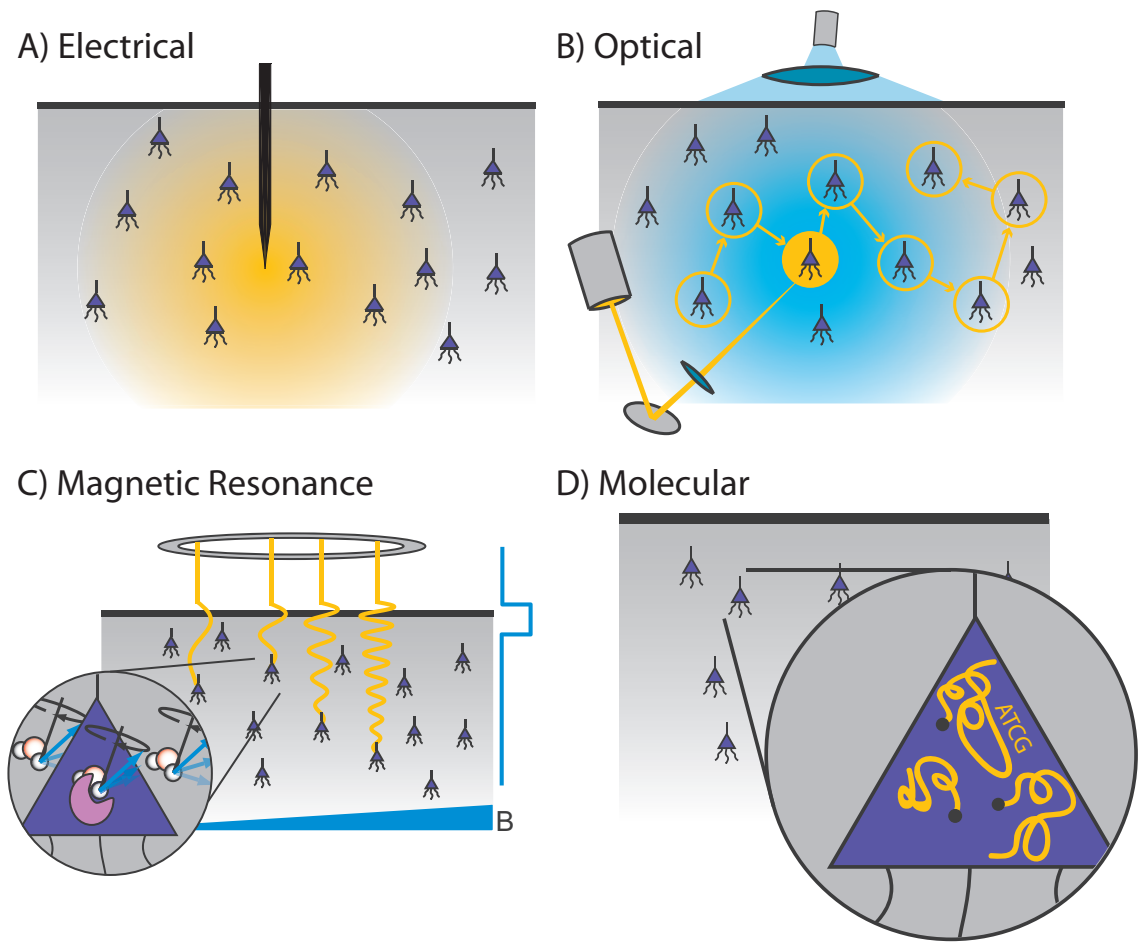

Our analysis is predicated on assumptions that enable us to estimate scaling limits. These include assumptions about basic properties of the brain, which are treated in section 2 (Basic Constraints), as well as those pertaining to the required measurement resolution and the limits to which a neural recording method may perturb brain tissue, which are treated in section 3 (Challenges for Brain Activity Mapping). Together, these considerations form the basis for our estimates of the prospects for scaling of neural recording technologies. We analyze four modalities of brain activity mapping — electrical, optical, magnetic resonance and molecular — in light of these assumptions, and conclude with a discussion on opportunities for new developments.

Importantly, our assumptions, analyses and the conclusions thereof are intended as *first approximations and are subject to debate*. We anticipate that as much can be learned from where our logic breaks down as from where it succeeds, and from methods to work around the limits imposed by our assumptions.

# 2  BASIC CONSTRAINTS

**Mouse brain**  The mouse brain contains $\sim$7.5 $\times 10^7$ neurons in a volume of $\sim$420 mm$^3$ [11] and weighs about 0.5 g. The packing density of neurons varies widely between brain regions. In the below, we will use a cell density of $\rho_{neurons} \approx 92\,000/mm^3$, as measured for mouse cortex [12]. This corresponds roughly to one neuron per 22 μm voxel. The density of cortical synapses, on the other hand, approaches $10^9/mm^3$, i.e., one synapse per 1 μm$^3$ voxel. For comparison, the human brain has roughly $8 \times 10^{10}$ neurons [13] in a volume of 1200 cm$^3$ [14].

The human brain consumes $\sim$15 W of power (performing, at synapses, a rough equivalent of at least $10^{17}$ floating point computational operations per second on that power budget, according to one definition [15], although the analogy with digital computers should not be taken literally). Because power consumption scales approximately linearly with the number of neurons [16], the mouse brain is expected to utilize $\sim$15 mW. For comparison, the metabolic rate of the $\sim$20–30 g mouse is $\sim$200–600 mW depending on its degree of physical activity [17].

**Neural activities**  Action potentials (spikes) last $\sim$2 ms. The rate of neuronal spiking is highly variable. Some authors have assumed an average rate of 5 Hz [15, 18], but certain neurons spike at 500 Hz or faster [19], while many neurons spike much more slowly. For example, cerebellar granule cells, which make up half of the neurons in the brain, have spontaneous firing rates of $\sim$0.5 Hz [20]. In neocortex, one estimate estimated 0.16 spikes per second per neuron (in primate) as energetically sustainable [21]. There may be as much as a two-fold change in metabolism and hence firing rate across brain states [22]. Certain neurons (possibly up to 90% for some neuron types in some brain areas) may be effectively silent [23, 24], e.g., spiking less than once every ten seconds. Some studies have attempted to measure the *distribution* of neural firing rates in various cortical areas (as opposed to just the average rate), and have observed that these distributions are often long-tailed: a small minority of the neurons fires a majority of the spikes [25–28].

While these estimates of typical firing rates are useful numbers to have in mind, in the below we aim to sample all neurons at $\sim$1 kHz rates (or higher for techniques requiring observation of detailed spike waveforms). This choice is informed by several factors. First, measuring spike *timing* with millisecond precision is relevant for understanding neural function, due to the possibilities for timing codes, spike-timing dependent plasticity mechanisms, and other effects relying on temporally-precise spiking patterns [29–32]. In this regard, it is also important for a recording method to maintain precise temporal phasing between measurements at different brain locations: activity measurements should be locked to precise global clocks, perhaps with a tolerable phase imprecision between any two measurements in the range of $\frac{1}{2\pi} \times 1$ ms $\approx$ 100–200 μs. Furthermore, the activities of neurons can be highly correlated locally or across large networks [33], suggesting that local activity sensors may be subjected to high instantaneous total firing rates due to simultaneously-active neurons.

**Absorption and scattering of radiation**  All existing methods of neural recording utilize electromagnetic waves, from the near-DC frequencies of wired electrical recordings ($\sim$1 kHz) to the radio-frequencies of wireless electronics and fMRI (MHz–GHz) to visible light in optical approaches ($\sim$500 THz). These electromagnetic waves are attenuated in brain tissue by absorption and scattering. As an approximation to the electromagnetic absorption by brain tissue, we treat the absorption by water, the brain's main constituent (68–80 % by mass in humans [36, 37]). At visible and near-IR wavelengths, scattering dominates absorption: absorption lengths are in the $\sim$1 mm range, while scattering lengths are $\sim$25–200 μm [38]. The combined effect of absorption and scattering is measured by the attenuation length, the distance over which the signal strength is reduced by a factor of $1/e$ along a path. Figure 2 shows the absorption length of water [35], and the attenuation length in a Mie scattering model (from [39]) intended to approximate the scattering properties of cortical tissue (and see [40] for tissue skin depth measurements in the 10 Hz to 100 GHz range). This gives a preliminary indication of which wavelengths can be used to measure deep-brain signals with external detectors. Note that the attenuation length is only one of several relevant metrics: for example, scattering not only causes signal attenuation, but also causes noise and impairs signal separation, so the magnitude of the scattering is a key figure of merit.

**Figure 2.** Penetration depth (attenuation length) of electromagnetic radiation in water vs. wavelength (data from [34]). The approximate diameter of the mouse brain is shown as a black dashed line. Inset: approximate tissue model based on Mie scattering theory and water absorption. Absorption length of water [35] (blue), approximate tissue scattering length in a simple Mie scattering model (red) and the resulting attenuation length (green) of infrared light (inset reproduced from [35], with permission).

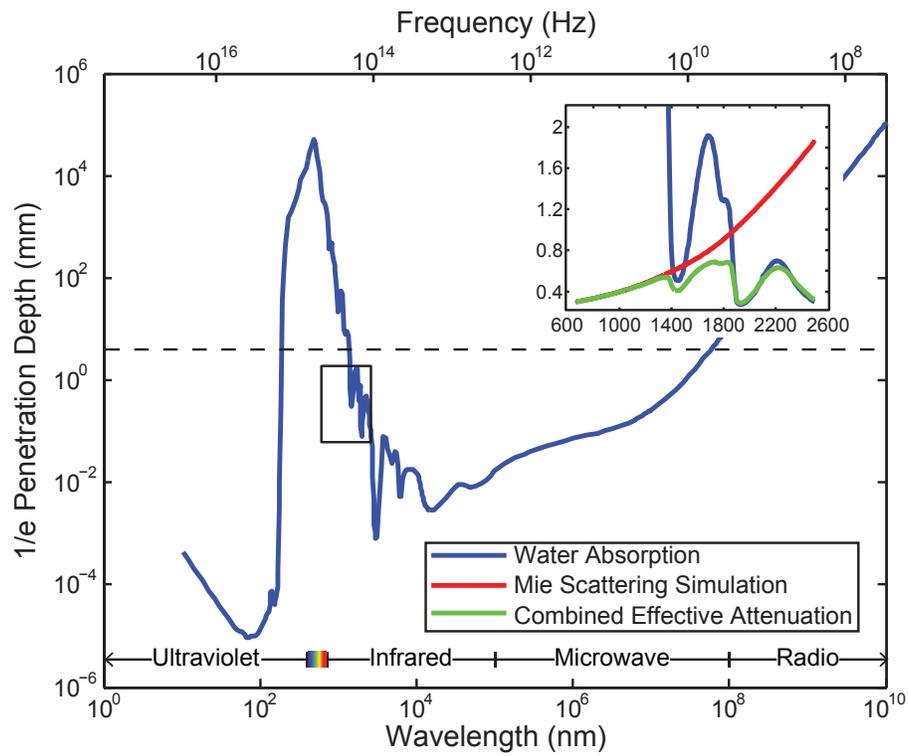

# 3  CHALLENGES FOR BRAIN ACTIVITY MAPPING

Any activity mapping technology must extract the required information without disrupting normal neuronal activity. As such, we consider three primary challenges: spatiotemporal resolution and informational throughput, energy dissipation and volume displacement.

## 3.1  SPATIOTEMPORAL RESOLUTION AND INFORMATIONAL THROUGHPUT

A sampling rate of 1 kHz is necessary to capture the fastest trains of action potentials at single-spike resolution. A minimal data rate of $7.5 \times 10^{10}$ bits processed per second is then required to record 1 bit per mouse neuron at 1 kHz.

In electrical recording, higher sampling rates (e.g. 10–40 kHz) are often necessary to distinguish neurons based on spike shapes when each electrode monitors multiple neurons. More fundamentally, one bit per neuron sampling at 1 kHz would likely not be sufficient to reliably distinguish spikes above noise: transmitting ~10 bit samples at ~10 kHz (full waveform) or ~10–20 bit time-stamps upon spike detection would be more realistic.

Conversely, it may be possible to locally compress measurements of a spike train before transmission. The degree of compressibility of neural activity data is related to the variability in the distribution of neural responses (e.g., such a distribution may be defined across time bins or repeated stimulus presentations) [41]. In the blowfly *Calliphora vicina*, the entropy of spike trains has been measured to be up to ~180 bit/s, and the information about a stimulus encoded by a spike train was as high as ~90 bit/s [41]. Extrapolating from fly to mouse, this would suggest that a compression factor of 5×–10× should be possible, relative to a 1000 bit/s raw binary sampling.

As a naïve estimate of the entropy as a function of firing rate, one can write the entropy $H$ in bit/s, assuming 1 ms long spikes and $f = 1000\,\mathrm{Hz}$ sampling rate, as

$$H \approx \left( -P_{\text{spike}} \cdot \log_2\left(P_{\text{spike}}\right) - \left(1 - P_{\text{spike}}\right) \cdot \log_2\left(1 - P_{\text{spike}}\right) \right) \cdot f$$

where $P_{\text{spike}}$ is the probability of spiking during the sampling interval (average firing rate/$f$). For an average firing rate of 5 Hz, $P_{\text{spike}} = 0.005$ and $H = 45\,\mathrm{bit/s}$, corresponding to a compression factor of ~20×. However, at 500 Hz average firing rate, $P_{\text{spike}} = 0.5$ with $H \approx 1000\,\mathrm{bit/s}$, i.e., there is no compressibility. Therefore, compression could conceivably reduce the data transmission burden for activity mapping by 1–2 orders of magnitude, depending on the neurons and activity regimes under consideration. Note that these compressibility calculations have assumed that firing patterns are independent across cells; they represent the temporal compressibility of the spike train from each cell, treated individually. Patterns across cells could conceivably be compressed by a much larger amount, to the extent that there is redundancy between cells. Nevertheless, we use 1 bit/neuron/ms or 100 Gbit/s as a "minimal whole brain data rate" in what follows. In many cases, this likely constitutes a lower bound on what is feasible in practice.

## 3.2  ENERGY DISSIPATION

Brain tissue can sustain local temperature increases ($\Delta T$) of ~2 °C without severe damage over a timescale of hours. Indeed, changes of this magnitude may occur naturally in rats in response to varying activity levels [42]. Assuming that the brain is receiving a constant power influx $P_{\text{delivered}}$ and that the local thermal transport properties of mouse brains are similar to those of humans, we can approximate the temperature change in deep-brain tissue as a function of the applied power [43, 44]:

$$\frac{dT}{dt} = \left( P_{\text{delivered}} + P_{\text{metabolic}} - \rho_{\text{blood}} C_{\text{blood}} f_{\text{blood}} \Delta T \right) \Big/ C_{\text{tissue}}$$

where $P_{\text{metabolic}} = 0.0116\,\mathrm{W/g}$ is the power per unit mass of basal metabolism, $C_{\text{tissue}} \approx 3.7\,\mathrm{J/(K\,g)} \approx 0.88 \cdot C_{\text{water}}$ is the specific heat capacity of brain tissue, $\rho_{\text{blood}} = 1.05\,\mathrm{g/cm^3}$ is the density of blood, $C_{\text{blood}} = 3.9\,\mathrm{J/(K\,g)}$ is the specific heat capacity of blood, $f_{\text{blood}} = 9.3 \times 10^{-9}\,\mathrm{m^3/(g\,s)}$ is the volume flow rate of blood, and $\Delta T$ is the temperature difference between the brain tissue and the blood (at 37 °C). A steady-state temperature increase ($dT/dt = 0$) of 2 °C corresponds to dissipation of ~40 mW per 500 mg mouse brain. Therefore, a recording technique should not dissipate more than ~40 mW of power in a mouse brain at steady state.

This estimate of the power dissipation limit in mouse brains, based on such a simplified model of the brain's thermal transport mechanisms, is likely an under-estimate of the actual maximum steady-state power dissipation. Radiative heat loss was ignored here since infrared light emitted by deep-brain tissue is quickly re-absorbed by nearby tissue. We have also ignored cooling due to flows in the cerebrospinal ventricles [45] and in the glymphatic system [46]. We have further assumed that conductive heat loss from the brain surface is negligible compared to the heat extracted volumetrically by blood flow. While this may hold true locally in deep brain voxels and over short timescales (e.g., < 1 min), further work (e.g., a whole-head model [44, 47]) is needed to define the true limits of sustained volumetric heat production by neural recording systems distributed throughout the mouse brain. Indeed, the characteristic length scale of temperature inhomogeneities in the brain is on the order of millimeters [48], whereas heat exchange with the flowing blood dampens the effects of local perturbations over longer length scales. For large brains, this means that sources and sinks of heat exert only local thermal effects; for a mouse brain on the scale of < 10 mm, however, surface and volumetric effects likely combine to influence temperature changes at any site in the brain [49]. Experimentally, increasing the temperature gradient at the brain surface, via a cranial window exposed to ambient air at ~25 °C (i.e., the common craniotomy technique used to access mouse neocortex), has been shown to dis-regulate brain temperature down to a depth of several millimeters [50]. For the above reasons, our estimates of the brain's capacity for heat dissipation should be treated only as first approximations.

Higher power levels, compared to the maximum steady state power, may be introduced into brains transiently. According to the above equation, if a neural recorder dissipates ~40 mW per 500 mg mouse brain, then the brain approaches the steady-state temperature in 2–3 min, making shorter experiments potentially feasible. This is in agreement with the estimate from [48] of a ~1 min time constant for brain temperature changes, as well as with experimental measurements showing similar time constants for temperature variations resulting from sustained neural stimulation [51, 52]. Increasing convective heat loss from the brain by increasing blood flow (e.g. via increased heart rate) or cooling the brain (volumetrically or via its surface) [49], the blood, the cerebrospinal fluid (CSF), or the whole animal [53], could increase the allowable transient or steady-state power dissipation.

There are also limits on the power density of radiation applied to brain tissue. For radio-frequency electromagnetic radiation, the specific absorption rate (SAR) limit on the power density exposed to human tissue is ~10 mW/cm² [54], while for ultrasound (which couples less strongly to dissipative loss mechanisms in tissue) the SAR limit is up to 72× higher [55]. The power density limit for visible and near-IR light exposures are also in the ~10–100 mW/cm² range for ~1 ms long exposures, decreasing as the exposure time lengthens (based on the IEC 60825 formulas [56]).

High local power dissipation (transient or steady-state) can modify the electrical properties of excitable membranes, altering neuronal activity patterns. For example, heating of cell membranes and of the surrounding solution by millisecond-long optical pulses leads to changes in membrane electrical capacitance mediated by the ionic double layer [57]. Slower temperature changes (on a scale of seconds) resulting from RF radiation lead to accelerated ion channel and transporter kinetics [58]. Both of these effects are appreciable when the temperature changes are on the order of 1–10 °C.

For comparison with current practice, common guidelines for chronic heat exposure from biomedical implants [42] use upper limits of 2 °C temperature change, 40 mW/cm² heat flux from the surface of implanted brain machine interface (BMI) hardware, and an SAR limit of

$$\frac{\sigma E^2}{2\rho} < 1.6 \, \text{mW/g}$$

for electromagnetic energy absorbed by tissue, where $E$ is the peak electric field amplitude of the applied radiation, $\sigma \approx 0.18 \, \text{S/m}$ is the electrical conductivity of grey matter and $\rho \approx 1 \, \text{g/cm}^3$ is the tissue density [44] (this corresponds to an irradiance of $\epsilon_0 c E^2 / 2 \approx 2.4 \, \text{mW/cm}^2$). A 96-channel BMI system demonstrated in living brains had dissipated areal power density approaching 40 mW/cm² [59].

## 3.3 Sensitivity to Volume Displacement

To prevent damage to the brain, we assume that a recording technique should not displace > 1 % of the brain's volume. *The appropriate damage threshold is not yet established, however, so this constitutes a first guess.* It is possible to insert large numbers of probes throughout multiple brain areas without compromising function. In rats, 96 electrodes of 50 μm diameter were simultaneously inserted across four forebrain structures (cortex, thalamus, hippocampus and putamen) [60]. In rhesus macaque, 704 electrodes of diameter 50 μm and average depth 2.5 mm were chronically implanted in cortex [61]. Note, however, that the total volume displacement in these experiments was below 0.1 %, and below 0.01 %, respectively. Furthermore, these studies used a low density of electrodes. Thus, detailed limits on the amount and density of inserted material are unknown.

Furthermore, the nature of the volume displacement is important — sheets of instrumentation that sever long-range connectivity, for example, would disrupt normal brain function regardless of the degree of volume displacement. Conversely, higher volume displacement might be possible if introduced gradually, or during early development, insomuch as the brain can adapt without disrupting natural computation. One important consideration in this regard would be the disruption of blood circulation by inserted material; a high density of implanted material in a brain region could cause stroke due to widespread vascular damage. Recent studies have defined in microscopic detail the complete vascular network of the mouse cortex using high-throughput histology [62]; this type of information could be used to enumerate key vascular pathways which could be spared from damage. To apply this in a particular animal, however, would require a non-destructive method to image the vasculature at a similar resolution; otherwise, only a broad statistical view can be obtained, since the detailed vascular geometry will vary from animal to animal.

Secondary effects like glial scarring may also pose obstacles to the long-term implantation of large numbers of probes [63, 64], although methods are being developed to alleviate this [65–67]. In the context of electrical recording, the impact of glial scarring may vary depending on geometry. For example, the recording sites at the tip of a Utah or Duke multi-electrode array are typically viable in chronic recordings of up to 18 months in primates [61, 68], whereas in array formats with multiple electrodes along each shaft, such as the Michigan array, chronic recordings of up to 4 months have been reported in rats [69]. Differences in recording lifetime may be due to differences in the pattern of glial encapsulation of the contacts.

# 4 Evaluation of Modalities

We next evaluate neural recording technologies with respect to the above challenges, using the mouse brain as a model system. Table 1 lists the modalities studied, the assumptions made, the analysis strategies applied, and the conclusions derived.

## 4.1 Electrical Recording

In the oldest strategy for neural recording, an electrode is used to measure the local voltage at a recording site, which conveys information about the spiking activity of one or more nearby neurons. The number of recording sites may be smaller than the number of neurons recorded since each recording site may detect signals from multiple neurons. As a note for practitioners, we use the term "electrode" interchangeably with the terms "recording site" or "contact", meaning a point-like voltage sensing node: many multi-electrode arrays in common use (e.g., the Duke and Utah arrays) are conductive only at the tip, whereas other designs (such as the Michigan array) have multiple contacts along the shaft. Each shaft in a Michigan array would thus constitute multiple "electrodes" or "recording sites" in our parlance. Traditional electrical recording techniques keep active devices such as amplifiers outside the skull and therefore do not pose a heat dissipation challenge; this may change if amplifiers are brought closer to the signal sources to reduce noise.

Slowly varying (e.g., $< 300\,\mathrm{Hz}$) extracellular potentials (LFPs) [71, 72] on the order of $0.1$–$1\,\mathrm{mV}$, and fields [73] on the order of $1$–$10\,\mathrm{mV/mm}$, are generated by neural activity. While LFPs can be filtered from the higher-frequency signals associated with extracellular voltage spikes, these and other effects necessitate maintaining precise potential references (i.e., ground levels) for voltage measurements distributed widely across the brain.

### 4.1.1 Spatiotemporal Resolution

**Limits assuming perfect spike sorting**     We begin with an idealized estimate of the number of electrodes required to record from the entire mouse brain, neglecting the difficulty of assigning observed spikes to specific cells (spike sorting), and focusing only on what is needed to detect spikes from every neuron on at least one electrode. The key variable here is the maximum distance between an extracellular electrical recorder and a neuron from which it records spikes. In a first approximation, this is determined by two factors: the decay of the signal with distance from the spiking neuron and the background noise level at the recording site. We assume that for an electrode to reliably detect the signal from a given neuron, the magnitude of that neuron's signal must be larger than the electrode's noise level. Note, however, that knowledge of spike shape distributions could potentially be used to extract low-amplitude spikes from noise.

The peak signals of spikes from neurons immediately adjacent to an electrode are in the $0.1$–$1.0\,\mathrm{mV}$ range and scale roughly as $e^{-r/r_0}$, where $r$ is the distance from the cell surface and the $1/e$ falloff distance, $r_0$, has been experimentally measured at $\sim 28\,\mu\mathrm{m}$ in both salamander retina [74] and cat cortex [75], and computed at $\sim 18\,\mu\mathrm{m}$ in a biophysically realistic simulation [76, 77]. However, this decay is strongly influenced by the detailed geometry of neuronal currents and the properties of the extracellular space (e.g., its inhomogeneity, which may lead to a frequency-dependent falloff of the extracellular potential [78]), making analytical calculation of the decay rate difficult (at large distances, a much slower $1/r^2$ dipole falloff is expected).

Several sources of background noise enter the recordings. Johnson noise, which arises from thermal fluctuations in the electrode, is

$$V_{\mathrm{johnson}} = (4k_{\mathrm{B}} T Z \mathrm{BW})^{1/2}$$

which for physiological temperature, electrodes of impedance $Z = 0.5\,\mathrm{M\Omega}$, and $\mathrm{BW} = 10\,\mathrm{kHz}$ bandwidth is $V_{\mathrm{johnson}} \approx 9\,\mu\mathrm{V}$. The recordings are also affected by interference from other neurons, which has been reported to exceed the Johnson noise, and is nonstationary due to changes in the cells' firing properties [79]. The noise and interference from these sources realistically produces $> 10$–$20\,\mu V$ of voltage fluctuations [80]. Current recording setups thus have signal to interference-plus-noise ratios (SINRs) of $< 100$, where the SINR is defined as the ratio of the peak voltage from immediately adjacent neurons to the voltage fluctuation floor of the electrode.

Table 1: Summary of modalities, models, assumptions and conclusions

| Modality | Analysis Strategy | Assumptions | Conclusions |
| --- | --- | --- | --- |
| *Extracellular electrical recording* | Compute minimal number of recorders based on max distance from recorder to recorded neuron<br><br>Compute channel capacity limits to spike sorting | Decay profile of extracellular voltage<br><br>Approximate noise levels at recording site | Maximum recording distance $r_{max} \approx 100$–$200\,\mu\mathrm{m}$ from electrode to neuron measured<br><br>$\sim 10^5$ recording sites are required per mouse brain at current noise levels assuming perfect spike sorting<br><br>$\sim 10^6$ recording sites are required at current noise levels at the physical limits of spike sorting<br><br>$\sim 10^7$ recording sites are required using current spike sorting algorithms |
| *Implanted electrical recorders* | Compute power dissipation of electronic devices that digitally sample neuronal activity | Physical limit: $k_{\mathrm{B}}T \ln(2)/\mathrm{bit}$ erased<br><br>Practical limit: $\sim 10 k_{\mathrm{B}}T/\mathrm{bit}$ processed<br><br>Current CMOS digital circuits: $> 10^5 k_{\mathrm{B}}T/\mathrm{bit}$ processed | Requires 2–3 orders of magnitude increase in the power efficiency of electronics relative to current devices to scale to whole-brain simultaneous recordings<br><br>Minimalist architectures could be developed to reduce local data processing overhead |
| *Wireless data transmission* | Compute tradeoff between power dissipation and channel bandwidth using information theory | Transmitter must supply enough power to overcome noise and path loss | Transmission at optical or near-optical frequencies is needed to achieve sufficient single-channel data rates using electromagnetic radiation. Radio-frequency (RF) electromagnetic transmission of whole-brain activity data draws excessive power due to bandwidth constraints<br><br>Bandwidth cannot be split over multiple independent RF channels, but IR light or ultrasound may allow spatial multiplexing |
| *Optical imaging* | Relate the scattering and absorption lengths of optical wavelengths in brain tissue to signal-to-noise ratios for optical imaging | Approximate values of scattering and absorption lengths as a function of wavelength | Light scattering imposes severe constraints, but strategies exist which could negate the effects of scattering, such as implantable optics, infrared indicators, signal modulation, and online inversion of the scattering matrix |
| *Multi-photon optical imaging* | Compute minimum total excitation light power to excite multi-photon transitions from indicators within each neuron in every imaging frame | Approximate values of multi-photon cross-sections<br><br>Pulse durations similar to those currently used in multi-photon imaging | Whole-brain multi-photon excitation will over-heat the brain except in very short experiments, unless ultra-high-cross-section indicators are used |
| *Beam scanning microscopies* | Calculate device and indicator parameters necessary for fast beam repositioning and signal detection | Fast optical phase modulators could reposition beams at $\sim 1\,\mathrm{GHz}$ switching rates<br><br>Fluorescence lifetimes in the 0.1–1.0 ns range | Beam repositioning time limits the speed of current systems but these are far from the physical limits<br><br>Fluorescence lifetimes of indicators constrain design of ultra-fast scanning microscopies |
| *Magnetic resonance imaging* | Calculate spatial and temporal resolution of MRI based on spin relaxation times and spin diffusion | Proton MRI using tissue water<br><br>Approximate $T_1$ and $T_2$ relaxation times and self-diffusion times for tissue water | Proton MRI is limited by the $T_1$ relaxation time of water to $\sim 100$ ms temporal resolution and by the self-diffusion of water to spatial resolutions of $\sim 40\,\mu\mathrm{m}$. $T_1$ pre-mapping could allow $T_2$ contrast on a $\sim 10$ ms timescale. Achieving these limits for functional imaging requires going beyond BOLD contrast |
| *Ultrasound* | Calculate spatial resolution, signal strength and bandwidth limits on ultrasound imaging | Speed of sound in brain<br><br>Attenuation length of ultrasound in brain | Attenuation of ultrasound by brain tissue and bone may be prohibitive at the $\sim 100$ MHz frequencies needed for single-cell resolution ultrasound imaging<br><br>Ultrasound may be viable for spatially multiplexed data transmission from embedded devices [70] |
| *Molecular recording* | Compute metabolic load and volume constraint for rapid synthesis of large nucleic acid polymers<br><br>Evaluate temporal resolution in simulated experiments using kinetic models [6] | Polymerase biochemical parameter ranges<br><br>Metabolic requirements of genome replication | Molecular recording devices appear to fall within physical limits but their development poses multiple major challenges in synthetic biology<br><br>Synchronization or time-stamping mechanisms are required for temporal resolution to approach the millisecond scale |

A limit on the maximum recording distance is the distance at which the signal from the farthest neuron falls below the noise floor, $r_{max} \approx r_0 \ln(SINR)$. For $SINR \approx 100$, $r_{max} \approx 130\,\mu m$. For comparison, recent experimental data from multi-site silicon probes has shown few detectable neurons beyond $\sim 100\,\mu m$ and none detectable beyond $160\,\mu m$ [81]. Recordings in the hippocampal CA1 region could not detect spikes from cells farther than $140\,\mu m$ from the electrode tip [82], even after averaging over observations triggered on an intracellularly recorded spike; in hippocampus, this corresponds to a detection volume containing approximately 1000 neurons [83]. Furthermore, in many studies (in monkeys, rats and mice) using multi-electrode arrays with 150–300 $\mu m$ inter-electrode spacings, no neuron is seen by more than one electrode [84–87].

Due to the steep local falloff, even improving the SINR by a factor of 10 only extends the maximal recording distance to $r_{max} \approx 190\,\mu m$. Assuming packing of the brain into equal sized cubes of side length $d = \frac{2\sqrt{3}}{3} r_{max} \approx 150\,\mu m$ gives $N > 130\,000$ electrodes for whole brain recording using recording sites with $r_{max} \approx 130\,\mu m$. Note that $N$ varies as the third power of $r_{max}$ and is therefore highly sensitive to variations in the assumed maximal recording distance; the number of required recorders can range from 38 000 to 210 000 as $r_{max}$ varies from $190\,\mu m$ to $110\,\mu m$.

These calculations, by assuming perfect spike sorting, greatly underestimate the required number of electrodes in practice. First, signals from the weakest cells are far weaker than those from the strongest cells and the signals from some cells decay much faster than others [75]. Second, because of neuronal synchronization, the local noise produced by nearby neurons may sometimes be large. Third, spike waveforms can vary over the course of a recording session [88, 89]. Finally, with many neurons per electrode or at high firing rates, spikes from detectable neurons will often temporally overlap, making spike sorting difficult.

**Figure 3.** The voltage signal to interference-plus-noise ratio (SINR) for neurons immediately adjacent to the recording site sets an approximate upper bound on the distance, $r_{max}$, between the recording site and the farthest neuron it can sense (blue), due to the exponential falloff of the voltage SINR with distance. Assuming at least one electrode per cube of edge length $\frac{2\sqrt{3}}{3} r_{max}$ in turn limits the number of neurons per recording site (gold), the total number of recording sites (red) and the maximal diameter of wiring consistent with $< 1\%$ total brain volume displacement (turquoise). SINR values for current recording setups are $< 10^2$. In practice, the number of neurons per electrode distinguishable by current spike sorting algorithms is only $\sim 10$, with an estimated information theoretic limit of $\sim 100$, so these curves *greatly under-estimate* the number of electrodes which would be required based on realistic spike sorting approaches in a pure voltage-sensing scenario.

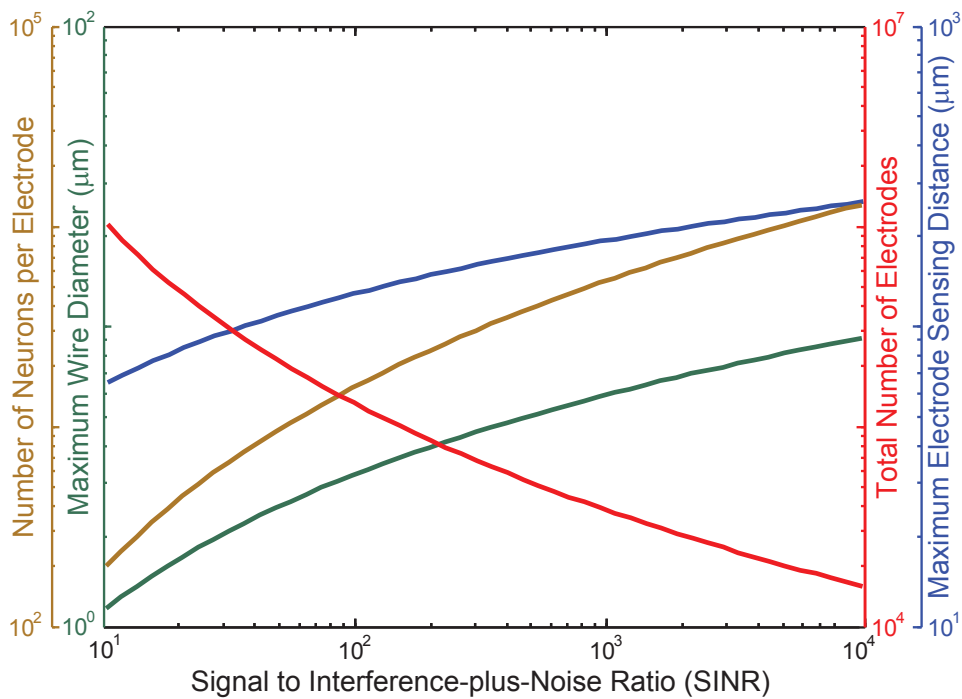

**Limits from spike sorting**     The previous calculations have assumed that any spike which is visible above the noise on at least one electrode can be detected and correctly assigned to a particular cell, i.e., that the problem of spike sorting can be solved perfectly. However, perfect spike sorting is far beyond current algorithmic capabilities and in fact may not be possible in principle.

To achieve the scenario described above, with $N = 130\,000$ recording sites per mouse brain, would require each electrode to sort spikes from all $\frac{4}{3}\pi r_{max}^3 \rho_{neurons}$ neurons in a sphere of radius $r_{max} \approx 130\,\mu m$ surrounding the recording site, where $\rho_{neurons} \approx 92\,000/mm^3$ is the density of neurons. This assigns $\sim 800$ neurons to a single electrode. Roughly half (i.e., 400) of these neurons will lie at $> 100\,\mu m$ distance from the electrode, and their signals on the electrode will therefore have voltage SINRs of $< 100e^{-100\,\mu m/28\,\mu m} \approx 2.8$, assuming as above that extracellular spike amplitudes decay exponentially in space.

Electrical recording can be viewed as a data transmission problem, with the electrode playing the role of a communication channel (see section 4.4). According to the Shannon Capacity Theorem [90], the information capacity $C$ of a single analog channel (with additive white Gaussian noise) is

$$C = BW \log_2(1 + S/N)$$

where BW is the bandwidth, $S$ is the signal power (proportional to the square of the voltage), and $N$ is the noise power. Here the bandwidth is $BW \approx 10\,kHz$, and the ratio of peak signal power to noise power of a single spike for the outer 400 cells is no more than $2.8^2$, or $0.5 \times 2.8^2$ using the RMS signal power instead of the peak. With 400 cells emitting $2\,ms$ spikes at $5\,Hz$, there will be an average of 4 cells spiking at a time, for $S/N \approx 0.5 \times 4 \times 2.8^2 \approx 15.7$ counting the signal power from all the spikes. The channel capacity is then $C \approx 40\,kbit/s$. This represents the maximum amount of information (e.g., about which neuron spiked when) that the population of spiking neurons can transmit via the electrode which measures them. To transmit *uniquely identifiable* signals from all 400 neurons at millisecond temporal precision, however, requires $1\,bit/s \times 400 = 400\,kbit/s$, which is $> 10\times$ greater than the channel capacity and is therefore not achievable. Even with optimal temporal compression of $\sim 5\,Hz$ spikes (see section 2), we would need to transmit $\sim 400/20 = 20\,kbit/s$, which is strictly less than the channel capacity and thus possible in principle, but barely so. Furthermore, the channel capacity given here is an overestimate, since 2.8 is an upper bound on the SINR of the outer cells. On the other hand, note that the use of a nominal $5\,Hz$ average firing rate here (in the estimates of signal to noise ratio and of temporal compressibility) greatly oversimplifies the distribution of firing rates across neurons, as discussed in section 2 above, so this analysis can only be treated as a first approximation.

Based on these rough estimates, perfect spike sorting may not be possible at $\sim 800$ neurons per electrode, in a sphere of radius $130\,\mu m$ surrounding a recording site, and at the noise levels typical of current electrodes. In essence, there may not be enough room on the electrode's voltage trace to discriminate such a large number of weak, noisy signals. Note that these information-theoretic limits still apply even if it is possible to resolve temporally overlapping spikes. In fact, the channel capacity is what ultimately limits the ability of a spike sorting algorithm to resolve such overlapping spikes.

To see the regime in which spike sorting becomes feasible, suppose that each electrode is only responsible for spike sorting from the population of $\sim 100$ neurons nearest to the electrode, i.e., in a sphere of radius $r \approx 64\,\mu m$, assuming the $92\,000/mm^3$ cell density from mouse cortex. The outermost 50% of these neurons are then positioned $> 50\,\mu m$ from the recording site. For these outermost 50 neurons, the voltage SINR is $< 100e^{-50\,\mu m/28\,\mu m} \approx 17$ and $S/N < 0.5 \times 17^2 \times (2\,ms \times 5\,Hz \times 50) \approx 72.3$. The channel capacity is therefore $< 62\,kbit/s$, whereas $50\,kbit/s$ is needed for signal transmission from 50 neurons without temporal compression versus $\sim 2.5\,kbit/s$ with temporal compression. Even 100 neurons per electrode may therefore still be close to the limits of information transmission through the noisy channel corresponding to a single electrode.

In practice these limits are likely to be highly optimistic, since the set of spikes emerging from a neuronal population is far from an optimally designed code from the perspective of multiplexed signal transmission through a voltage-sensing electrode: the waveforms for different neurons are similarly-shaped rather than orthogonal, the spikes emitted by a given neuron vary somewhat in amplitude and exhibit shape fluctuations (signal-dependent noise), and it is not known in advance what the characteristic signal from each neuron looks like (or even how many neurons there are).

Indeed, current practice is far from the above information-theoretic limits. At present, spike sorting algorithms operating on data from large-scale (250-500 electrodes), densely spaced ($\sim 30\,\mu m$), 2D multi-electrode arrays can reliably identify and distinguish spikes from nearly all of the 200-300 retinal ganglion cells [91, 92] in a small patch of retina, and can also infer approximate cell locations through spatial triangulation of spike amplitudes. This represents a roughly $1:1$ ratio of cells to electrodes. Electrodes with up to 4 single units can be found in chronically implanted multi-electrode arrays (in both mouse and primate) [61, 93], where the electrodes are sparse, although the average yield of cells per electrode is closer to $1:1$; if only electrodes with at least one cell are counted, the average rises to $\sim 1.5$–$1.7$ cells per electrode. Optimistically, simulations of neural activity suggest that 5-10 neurons per electrode may be distinguishable using current spike sorting algorithms [79, 80, 94]. A limit of $\sim 10$ neurons per electrode would imply $N = 7.5 \times 10^6$ electrodes to record from all neurons in the mouse brain, which could be accomplished by positioning recording sites on a cubic lattice with $\sim 40\,\mu m$ edge length.

Future algorithmic improvements could enable sorting from more than ∼10 cells per electrode, but this becomes increasingly challenging. One simple estimate of a reasonable practical limit, for the regime of many neurons per electrode, would be the largest number of neurons that can be sorted without requiring the frequent resolving of temporally overlapping spikes: if the average neuron fires at ∼5 Hz and spikes last ∼2 ms, then at most roughly 100 neurons per electrode can be sorted without requiring overlaps to be resolved. Note that while some present-day algorithms can successfully resolve overlapping spikes [74, 91, 92, 95, 96], they typically do so only in the case where electrodes are densely spaced and any given spike appears on many electrodes, such that spatial information can be used to resolve the overlap. Resolving overlaps when spikes appear on only one or a few channels is more difficult due to noise and spike-shape variation.

Overall, ∼100 cells per electrode may be taken as a rough estimate of the limits of spike sorting, and would imply $N = 750\,000$ electrodes and an edge spacing of ∼80 μm if a cubic lattice of recording sites were used. However, we should not exclude the possibility of game-changers which could alter the nature of the recorded data to improve the available information. For instance, CCD cameras could be attached to multi-electrode arrays to aid in the identification and localization of cells, or directional information on the source of spikes could be obtained at each recording site, for example by measuring the directions of gradients in voltage. Systems that capture such additional information could circumvent the above information-theoretic limits and improve spike sorting.

### 4.1.2 VOLUME DISPLACEMENT

We require < 1 % total volume displacement from $N$ recorders. Wires from each electrode must make it to the surface of the brain, which implies an average length $\ell \approx 4$ mm for the mouse brain (depending on assumptions about the wiring geometry).

As a rough approximation, consider each recorder to produce a volume displacement associated with a single cylindrical wire, with length $\ell$ and radius $r$. Thus $r$ must satisfy

$$\pi r^2 \ell N_{min,rd} < 0.01 V_{brain}$$

Using $N_{min,rd} = 210\,000$ or 38\,000 recording sites (lower and upper limits from the perfect spike sorting case from above) and $\ell \approx 4$ mm requires wires of radius $r_{max} \approx 6.0$ μm, or 2.5 μm, respectively. Alternatively, if $7.5 \times 10^6$ electrodes must be used (current spike sorting case from above), the required wire radius is ∼200 nm. While these dimensions are readily achievable using lithographic fabrication, there would be a challenge to produce *isolated* wires of such dimensions at scale (perhaps suggesting the use of wire bundles). Still, volume constraints per se are unlikely to fundamentally limit whole-mouse-brain electrical recording even in the most pessimistic scenario.

Figure 3 illustrates the above considerations as a function of the electrode SINR.

### 4.1.3 IMPLANTING ELECTRODES IN THE BRAIN

There are several technology options for introducing many electrodes into a brain. For example, flexible nanowire electrodes could, in theory, be threaded through the capillary network [97]. Capillaries are present in the brain at a density of 2500–3000 per mm³ [98], which equates to one capillary per 73 μm, with each neuron lying within ∼200 μm of a capillary [99]. The minimum capillary diameter is as small as 3–4 μm, although the average diameter is ∼8 μm, comparable to the non-deformed size of the red blood cells [100]. Blocking a significant fraction of capillaries could lead to stroke or to unacceptable levels of tissue necrosis/liquifaction.

The cerebrospinal ventricles may also provide a convenient location for recording hardware. Furthermore, neural tissues could be grown around pre-fabricated electrode arrays [101], or silicon probes arrays with many nano-fabricated recording sites per probe [81] could be inserted into the brain.

Mechanical forces during insertion and retraction of silicon and tungsten microelectrodes from brain tissue have been measured in rat cortex at ∼1 mN for electrodes of ∼25 μm radius [102]. These forces are comparable to the Euler buckling force $F$ of a 2 mm long cylindrical tungsten rod of $r = 5$ μm radius

$$F = \frac{\pi^2 E I}{(KL)^2} \approx 1 \text{ mN}$$

where $E = 411$ GPa is the elastic modulus of tungsten, $I = (\pi/2) r^4$ is the moment of inertia of the wire cross-section, $L \approx 2$ mm is the length of the wire, and $K$ is the column effective length factor which depends on the boundary conditions and is set to $K = 1$ here for simplicity. This suggests that it may be possible to push structures of < 10 μm diameter into brain tissue (see [103] for related calculations). It might be advantageous to pull rather than push wires into the brain (e.g., using applied fields, or perhaps even cellular oxen [104] to carry the wires), since the thinnest wires could withstand tension forces much higher than the compressive force at which they buckle (although there may also be ways to circumvent buckling, e.g., via rapid vibration).

#### 4.1.4 Conclusions and Future Directions

Electrical recording has the advantage of high temporal resolution, but the large number of required recording sites poses challenges for delivery mechanisms. Ongoing innovations in electrical recording that could be leveraged for dramatic scaling include the development of highly multiplexed probes, multilayer lithography for routing electrical traces, novel methods to implant large numbers of electrodes, smaller electrode impedances to reduce the Johnson noise, amplifiers with lower input-referred noise levels, spike sorting algorithms capable of handling temporally overlapping spikes and adaptively modeling the noise, and hybrid systems integrating electrical recording with implantable optics or other methods.

One challenge for a purely-electrical recording paradigm pertains to the ability to relate the measured electrical signals to specific cells within a circuit. As the set of neurons recorded by each electrode grows to encompass a large volume around the electrode, it will become more difficult to attribute the recorded spikes to particular neurons. Furthermore, given the complex geometries of neuronal processes, it is not obvious how to determine the spatial position or layout of a neuron from its electrical signature on a nearby electrode. A given electrode will be positioned near the axons or dendrites of some neurons, and near the cell bodies of other neurons, complicating data interpretation. If the spatial density of recording sites is increased such that many electrodes sample the same neuron, however, this could enable imaging of neuronal morphology and signal propagation via voltage signals across multiple electrodes [105]. Currently, extracellular electrical recording also does not allow extraction of molecular information on the cells being recorded, although intracellular electrophysiological recording methods (e.g., [106]) might enable this for a limited number of cells.

### 4.2 Optical Recording

Optical techniques measure activity-dependent light emissions from neurons, typically generated by fluorescent indicator proteins, although activity-dependent bioluminescence emissions are an emerging possibility. Current genetically encoded calcium indicators can only distinguish spikes below ~50–100 Hz firing rates without averaging [107] due to slow intra-molecular kinetics and indicator saturation at high firing rates, although significant improvements in speed are ongoing [108]. Intracellular calcium rises and drops can occur within 1 ms and 10–100 ms respectively [109], which sets the ultimate speed limit for calcium imaging. The field of genetically-encoded high-speed fluorescent voltage indicators is also advancing quickly [110–115] and these may find particular use in monitoring sub-threshold events [116].

#### 4.2.1 Spatiotemporal Resolution

**Multiplexing strategies**  For optical approaches, the light originating from the activity of each neuron must be separated from emissions originating from other points in the brain: this can be accomplished in many ways, leading to a variety of architectures for 3D imaging. *Epi-fluorescence microscopy* images a plane in the specimen (i.e., with depth of field DOF $= \frac{2n\lambda}{NA^2}$, where $n$ is the refractive index, $\lambda$ is the wavelength and $NA$ is the numerical aperture of the imaging system [117]) onto a spatially-resolved two-dimensional detector (e.g., a CCD camera). The focal plane is then scanned in order to reconstruct 3D images; because the entire 3D volume is illuminated during image acquisition, out-of-focus neurons cause background emissions. *Light sheet imaging* is similar to epi-flourescence imaging, except that only neurons near the focal plane are illuminated, reducing out of focus noise. Unfortunately, this requires transparent brains [8]. Volumetric imaging can also be performed in a single snapshot using *lightfield microscopes* [118, 119], which capture the directions of incoming light rays, trading in-plane resolution for axial resolution, or by using multi-focus microscopes [120]. In *multi-photon microscopy*, nonlinearities result in fluorescence excitation occurring only near the focal point of the excitation laser, which is scanned across the sample. In *confocal scanning microscopy*, only photons from a point of interest are measured due to geometric constraints (e.g., pinholes). Alternatively, 3D imaging can be performed via *wavefront coding*, which extends the depth of field by creating an axially-independent point-spread function using known optical aberrations, in combination with computational deconvolution [121]. With a known 3D pattern of excitation light, wavefront coding can be applied to 3D fluorescence microscopy without scanning using a 2D detector array [117]. Emerging, alternative strategies rely on *tagging* emissions from different sources with distinguishable modulation patterns [122–126], or precisely controlling and tracking the timing of light emissions [127]. Optical techniques thus achieve signal separation by multiplexing spatially (e.g., direct imaging) or temporally (e.g., beam scanning), or often by a combination of the two.

While optics might seem to require a number of photodetectors comparable to the number of neurons (or a similar number of sampling events in the time domain, e.g., for scanning microscopies), new developments suggest ways of imaging with fewer elements. For example, compressive sensing or ghost imaging techniques based on random mask projections [128–131] might allow a smaller number of photodetectors to be used. In an illustrative case, an imaging system may be constructed simply from a single photodetector and a transmissive LCD screen presenting a series of random binary mask patterns [132], where the number of required mask patterns is much smaller than the number of image pixels due to a compressive reconstruction.

**Effects of light scattering**     Single-photon techniques limit imaging to a depth of a few scattering lengths at the excitation and emission wavelengths of activity indicators: up to ~1–2 mm for certain infrared wavelengths [39, 133, 134] vs. a few hundred microns for visible wavelengths [38]. Activity dependent dyes are currently available only in the visible spectrum; indicators operating in the infrared (see [135–137] for far-red fluorescent proteins) could improve imaging depth.

Multi-photon excitation takes advantage of the deeper penetration of infrared light. Two or more infrared photons may together excite a fluorophore with an excitation peak in the visible range, leading to the emission of a visible photon. If only one neuron is illuminated with sufficient intensity to generate multi-photon excitation, all photons captured by the detector originate from that neuron, regardless of the scattering of the outgoing light. Hence, the emission pathway is limited less by scattering than by absorption. This has resulted in imaging at > 1 mm depth [39, 133, 134].

There are at least five options for overcoming visible light scattering to enable signal separation from deep-brain neurons [1, 138]:

1. Infrared light can excite multi-photon fluorescence in an excitation-scanning architecture.

2. Fluorophores with both excitation and emission wavelengths in the infrared could be developed.

3. By knowing the precise form of the scattering, it can be possible to correct for it. Emerging techniques based on beam shaping allow transmission of focused light through random scattering media by inverting the scattering matrix [139]. Because the scattering properties change over time, this must be done quickly, possibly faster than the imaging frame rate, necessitating high-speed wavefront modulation. This can currently be achieved with digital micro-mirror devices (DMDs), but not with the phase-only spatial light modulators (SLMs) that are used to prevent power losses in the excitation pathways for nonlinear microscopies, although GHz switching of phase-only modulators appears feasible in principle [138]. High speed focusing through turbid media is also achievable using all-optical feedback in a laser cavity [140], and it is even possible to measure the scattering matrix non-invasively [141] using a photo-acoustic technique, or via all-optical approaches based on speckle correlation [142]. Similar techniques are available for incoherent light [143].

   When using short optical pulses, scattering can lead to temporal distortions that degrade the peak light intensity at a focal spot. The < 100 fs pulse durations used in two-photon microscopy, for example, are comparable to the time it takes light to travel 30 μm in vacuum. Fortunately, wavefront shaping techniques can correct for scattering-induced temporal distortions as well [144, 145].

4. Light sources and/or detectors could be positioned close to the measured neurons, necessitating the use of embedded optical devices. This could be done using optical fiber [146] and/or waveguide [147, 148] technologies, which are developing rapidly. For example, single-mode fiber cables can support > 1 TB/s data rates [149, 150] with low light loss over hundreds of kilometers [151]. It is possible to directly image through gradient index of refraction (GRIN) lenses [152] or optical fibers [146, 153, 154], which provides one way to multiplex many observed neurons per fiber.

5. Light emissions from distinct locations can be tagged with distinguishable time-domain modulation patterns, and the emission time-series for each source can later be decoded from the summed signal resulting from scattering [122–127]. For example, ultrasound encoding [122], which frequency-tags light emissions from a known location via a mechanical Doppler shift of the emitter [155], provides a generic mechanism to sidestep problems of elastic optical scattering, although it requires distinguishing MHz frequency modulations in THz light waves (part per million frequency discrimination). Radio-frequency tagging of light emissions via a digitally synthesized optical approach is also an option and may be applicable to combatting the problem of emission scattering in deep-tissue, multi-point, multi-photon imaging [123].

**Speed of beam scanning**     The speed of scanning microscopes is currently limited by beam repositioning times (~0.1 μs for spinning disk [146, 153, 154], ~3 μs for piezo-controlled linear scan mirrors, ~10 μs for acousto-optic deflectors [156], ~8 kHz line scans for resonant galvanometer mirrors). The 10 μs repositioning time for acousto-optic deflectors is set by the speed of sound in the deflector crystal, while scanning mirrors and spinning disks are limited by inertia. Note that 0.1 μs repositioning time for current spinning-disk confocal techniques would require 10 seconds per frame for whole mouse brain imaging with a single scanned beam $\left(10^{-7}\,\text{s/site} \times 10^{8}\,\text{sites/brain}\right)$. There is therefore a need for a $10^4$ fold improvement in beam repositioning time and/or beam parallelization in order to achieve 1 kHz imaging frame rates for whole mouse brains.

One strategy to implement parallelization would exploit (yet to be developed) fast, high-resolution phase modulator arrays to arbitrarily re-shape coherent optical wavefronts for multisite holographic multi-photon excitation in 3D [138, 157, 158]. With fast phase modulation (e.g., ~1 GHz), beating each excitation spot at a different frequency could allow a single detector to probe multiple sites in parallel, despite arbitrarily-large scattering of the outgoing light [138]. Emerging optical techniques may provide alternative means to implement similar strategies [123]. Temporal multiplexing of excitation pulses at distinct locations (e.g., via few-nanosecond beam delays) also allows parallelization of the excitation beam while combating scattering ambiguity of the emitted light [127]. Furthermore, temporal focusing techniques in two-photon microscopy (depth-dependent pulse duration) can excite an entire plane or line within the sample [159–162], as well as arbitrary patterns of points [157], potentially allowing fast axial scanning (somewhat analogous to light-sheet techniques used with transparent samples). This method intrinsically corrects for scattering of the excitation light [163], although not of the emission light. Like other multi-photon techniques, however, all these methods remain highly dissipative, as discussed below.

Fluorescence lifetimes in the 0.1–1 ns range [164] ultimately constrain the design of scanning fluorescence microscopies. A delay of 0.1 ns per mouse neuron per frame corresponds to only 100 Hz frame rate without parallelization, implying that parallelization into at least 10 to 100 beams is essential. The fluorescence lifetime also limits the achievable modulation frequencies in beat-frequency-multiplexed parallelization strategies [123], bit lengths in encoded strategies [124], and temporal offsets in temporally-multiplexed strategies [127], suggesting that parallelization of detectors may be necessary in a strongly scattering environment. Depending on the degree of parallelization, which constrains the achievable dwell times given a fixed frame rate, photon counts may also become a limiting factor for high-speed scanning in some approaches.

**Diffraction**     Using the small angle approximation, the diffraction-limited angular resolution of an aperture is $\theta \approx \frac{\Delta x}{y} \approx \frac{\lambda}{D}$, where $\Delta x$ is the spacing which must be resolved, $y$ is the imaging depth, $\lambda$ is the wavelength, and $D$ is the aperture diameter. Thus distinguishing neurons which are 10 μm apart and at a depth of 10 mm requires a lens aperture $D$ of $> 1$ mm when $\lambda \approx 1$ μm. Diffraction therefore does not appear to be a limiting factor for cellular resolution imaging, except in the context of microscale apertures that might find use in embedded optics approaches.

### 4.2.2 ENERGY DISSIPATION

Light that does not leave the brain is ultimately dissipated as heat. The total light power requirements for optical measurement of neuronal activity using fluorescent indicators depend on factors including fluorophore quantum efficiency, absorption cross-section, activity-dependent change in fluorescence, background fluorescence, labeling density, activation kinetics, detector noise, scattering and absorption lengths, and others. Unfortunately, many of these variables are unknown or highly dependent on particular experimental parameters.

A statistical analysis of photon count requirements for spike detection (in the context of calcium imaging) can be found in [165], which derived a relationship between the number of background photon counts ($N_{bg}$) and the number of signal photon counts required for high fidelity spike detection given photon shot noise. This scales roughly as $N_{signal} > 3\sqrt{2N_{bg}}$, even at low absolute photon count rates. While this analysis governs the number of detected photons, the number of emitted photons will be higher due to losses. In one example using two-photon excitation, 5 % of the emitted photons were captured by the photodetector [166]. One implication of photon shot noise is that faster-responding indicators (e.g., voltage indicators which respond in near-real-time to the membrane potential) must be brighter.

**Multi-photon excitation**     Multi-photon experiments rely on short laser pulses with high peak light intensities at a focused excitation spot to excite nonlinear transitions [166]. This imposes an experimentally relevant physical limit: at least one excitation pulse of sufficient intensity per neuron per frame is required in order to excite multi-photon fluorescence during each frame. Assuming 1 kHz frame rate and 0.1 nJ pulses [127], delivering only one pulse per neuron per frame would dissipate roughly $\left(10^8 \times 1\,\text{kHz} \times 0.1\,\text{nJ}\right) 10\,\text{W}$ in the mouse brain, which is clearly prohibitive. This is a lower bound because, in general, more than one excitation pulse per neuron per frame may be required to excite detectable fluorescence (e.g., one reference reported 12 pulses per spot [166]). For three-photon excitation, the situation will be even worse as higher peak light intensities are required to excite three-photon fluorescence.

Could the single-pulse energy be reduced while maintaining efficient two-photon excitation? The number of two-photon (2P) transitions excited per fluorophore per pulse is $n_a = F^2 C / t$, where $F$ is the number of photons per pulse per area in units of photon/cm$^2$, $C$ is the two-photon cross-section in units of cm$^4$s/photon, and $t$ is the pulse duration in seconds. This can be approximated as

$$n_a = \left( \frac{\frac{E}{hc/\lambda}}{\left(\frac{\lambda}{2(\text{NA})}\right)^2} \right)^2 \frac{C}{t} = \left( \frac{4E\,(\text{NA})^2}{hc\,\lambda} \right)^2 \frac{C}{t}$$

where NA is the numerical aperture of the focusing optics, $E$ is the pulse energy and $\lambda$ is the stimulation wavelength. For a 2P experiment with 100 fs, 0.1 nJ pulses, assuming a 2P cross section [167, 168] of $10^{-48}$ cm$^4$s/photon (i.e., 100 Goeppert-Mayer units [169], comparable to that of DsRed2 [168]), $\lambda = 900$ nm and NA = 1.0, $n_a \approx \frac{1}{2}$. Thus, a few pulses are likely necessary and sufficient to excite 2P fluorescence by each fluorophore within the focal spot. With a 2P cross section above $10^{-47}$ cm$^4$s/photon (1000 Goeppert-Mayer units, higher than that of any fluorescent protein that we are aware of [168]), one could reduce the pulse energy by an order of magnitude (and hence $n_a$ by two orders of magnitude) while maintaining $n_a > \frac{1}{20}$, i.e., one in twenty fluorophores excited by each pulse. Reducing the pulse energy much further might lead to unacceptably low excitation levels. Alternatively, shorter pulse durations could increase the light intensity, and hence 2P excitation probability, at fixed pulse energy.

Quantum dots can have 2P cross sections much higher than those of fluorescent proteins: water-soluble cadmium selenide–zinc sulfide quantum dots have been reported with 2P cross sections of 47000 Goeppert-Mayer units and are compatible with in-vivo imaging [170]. These would allow excitation efficiencies of $n_a \simeq \frac{1}{20}$ at pJ pulse energies, bringing whole-brain 2P imaging into the $\sim 100\,\mathrm{mW}$ range. Thus, the use of quantum dots or other ultra-bright multi-photon indicators could be decisive for supporting the energetic feasibility of multi-photon methods at whole brain scale; there are also plausible strategies for coupling quantum dot fluorescence to neuronal voltage [171]. However, some quantum dots have long fluorescence lifetimes [172], which may constrain scan speed.

For comparison to current practice, in a typical multi-photon experiment on mice, $\sim 50\,\mathrm{mW}$ of time-averaged laser power at the sample was used with a dwell time of $\sim 3\,\mu s$ [173], corresponding to $\sim 150\,\mathrm{nJ}$ energy dissipation per spot per frame. This dwell time would allow imaging only $\sim 300$ neurons at millisecond resolution with a single scanned excitation beam. The average excitation power here is likely already close both to whole-brain thermal dissipation limits, and to photo-damage limits for pulsed two-photon excitation [174, 175].

### 4.2.3 Bioluminescence

To work around the requirement for large amounts of excitation light, bioluminescent rather than fluorescent activity indicators could be used [176–178]. Consider a hypothetical activity-dependent bioluminescent indicator emitting at $\sim 1700\,\mathrm{nm}$ (IR), in order to evade light scattering. As a crude estimate, assuming that 100 photons must be collected by the detector per neuron per 1 ms frame, and 1 % light collection efficiency by the detector relative to the emitted photons, $\sim 100\,\mu\mathrm{W}$ of bioluminescent photons emissions are required for the entire mouse brain (using $E_{photon} = hc/\lambda$). This would be feasible from the perspective of heat dissipation. By contrast, in a 1-photon fluorescent scenario, if 100 excitation photons must be delivered into the brain to generate a single fluorescent emission photon, the power requirement becomes $10\,\mathrm{mW}$, which is on the threshold of the steady-state heat dissipation limit. Therefore, bioluminescent indicators could potentially circumvent problems of heat dissipation even in the 1-photon case.

The widely used bioluminescent protein firefly luciferase is $\sim 80\,\%$ efficient in converting ATP hydrolysis coupled with luciferin oxidation into photon production, yielding $\sim 0.8$ photons per ATP-luciferin pair consumed [179], and has $\sim 90\,\%$ energetic efficiency in converting free energy to light production. Heat dissipation associated with the luciferase biochemistry itself is therefore not a significant overhead relative to the $100\,\mu\mathrm{W}$ of emitted photons calculated above. In the same scenario, however, each neuron would consume $\sim 6 \times 10^8$ additional ATP molecules per minute in order to power the bioluminescence, which is within the limits of cellular aerobic respiration rates ($\sim 1\,\mathrm{fmol}\ O_2$ per minute per cell [180], with $\sim 30$ ATP per 6 $O_2$, hence $3 \times 10^9$ molecules ATP synthesized per minute from ADP via glucose oxidation), but not by a large margin. Transient increases in metabolic rate are possible: energy dissipation more than doubles in the mouse during high physical activity [17]. Therefore, whole-brain activity-dependent bioluminescence, at speeds high enough to achieve millisecond frame rates, may be metabolically taxing for the cell but is nevertheless plausible as a light generation strategy. Note that we have not treated the energy required to bio-synthesize the luciferin compound, which may create additional overhead (though conceivably luciferin could be provided exogenously).

### 4.2.4 Conclusions and Future Directions

Scattering of visible light in the brain creates a problem of signal-separation from deep-brain neurons. Multi-photon techniques, which scan an infrared excitation beam, can work around this scattering problem. However, current multi-photon techniques using fluorescent protein indicators, when applied at whole brain scale, would dissipate too much power to avoid thermal damage to brain tissue. Systems (such as plasmonic nano-antennas [181] or subwavelength metallic gratings [182]) that could locally excite multi-photon fluorescence without the need for high-energy laser pulses could conceivably ameliorate this issue. Importantly, quantum dots show promise as ultra-bright multi-photon indicators, if they can be targeted to neurons and optimized in terms of fluorescence lifetime. New methods besides multi-photon techniques could also work around the scattering of visible light in the brain. For example, fluorophores or bio-luminescent proteins could be developed which operate at infrared wavelengths. A compelling example from nature is the black dragonfish, which generates far red light ($\sim 705\,\mathrm{nm}$) via a multi-step bioluminescent process (using this light to see in deep ocean waters) [183, 184]. A large set of activity indicators with distinguishable colors, generated through a combinatorial genetic recombination mechanism such as BrainBow [185], could also improve signal separation. Targeting, via protein tags, of activity indicators to specific locations — such as the axon, soma, soma and proximal dendrites, distal dendrites, pre-synaptic terminals, post-synaptic terminals, or intact synapses — could also aid in signal discrimination [186–193]. In addition, implanted optical devices, which place emitters and detectors within a few scattering lengths of the neurons being probed, could potentially obviate the negative effects of scattering and allow visible-wavelength indicators to be used without a need for multi-photon excitation. In principle, excitation and detection do not need to make use of the same modality. For example, photoacoustic microscopy [194] uses pulsed laser excitation to drive ultrasonic emission, leading to optical absorption contrast. Such asymmetric techniques impose fundamentally different requirements from pure-optical techniques relative to fluorophore properties, required light intensities and other parameters.

## 4.3 EMBEDDED ACTIVE ELECTRONICS

The preceding sections have assumed that electrical or optical signals from the recorded neurons are shuttled out of the brain before digitization and storage, but it is also conceivable to develop embedded electronic systems that locally digitize and then store or transmit (e.g., wirelessly) measurements of the activities of nearby neurons. This could allow for shorter wires in electrical recording approaches, and for shorter light path lengths in optical recording approaches, as well as for more facile (e.g., non-surgical) delivery mechanisms for the recording hardware.

Integrated circuits have shrunk to a remarkable degree: in about 3 years, following the Moore's law trajectory, it will likely be possible to fit the equivalent of Intel's original 4004 micro-processor in a $10\,\mu m \times 10\,\mu m$ chip area. Functional wirelessly powered radio-frequency identification (RFID) chips as small as $50\,\mu m$ in diameter have been developed [195] and tags with chip-integrated antennas function at the $400\,\mu m$ scale [196]. Integrated neural sensors including analog front ends are also scaling to unprecedented form factors: a $250\,\mu m \times 450\,\mu m$ wireless implant – including the antenna, but not including a $\sim 1\,mm$ electrode shank used to separate signal from ground – draws only $2.5\,\mu W$ per recording channel [197]. The system operates at $\sim 1\,mm$ range in air, powered by a transmitter generating $\sim 50\,mW$ of transmitted power. Note that for a single such embedded recording device, the heat dissipation constraint is set not by the device's own dissipation ($10\,\mu W$ for four recording channels) but rather by the RF specific absorption rate limit associated with the $50\,mW$ transmit power.

Possibilities may exist for non-surgical delivery of embedded electronics to the brain: remarkably, cells such as macrophages ($\sim 13\,\mu m$ in size) can engulf structures up to at least $20\,\mu m$ in diameter [198] and have been studied as potential delivery vehicles for nano-particle drugs [199], suggesting that they might be used to deliver tiny microchips. T-cells and other immune cells can trans-migrate across the blood brain barrier [200] and ghost cells (membranes purged of their contents) engineered to encapsulate synthetic cargo [201] can fuse with neurons [202]. It might even be possible to engineer such cell-based delivery vehicles to form electrical gap junctions [203] with neurons or to act as local biochemical sensors [204].

The real-time transmission bandwidth requirements for neural recording could be significantly reduced if it is only desired to take a "snapshot" of neural activity patterns over a limited period of time, but this would require a large amount of local storage. For example, flash memory can store $> 10\,Mbit$ of data in a device $100\,\mu m$ on a side: a 64 giga-byte microSD card with $1.5\,cm^2$ area corresponds to 34 mega-bits per $(100\,\mu m)^2$ area. Even denser forms of memory storage are under development and could perhaps be used in a one-time-write mode in the context of neural recording long before they become commercially viable for use as rewritable media in the electronics industry.

Here we consider the power dissipation associated with embedded electronic recording devices, as well as the constraints on possible methods to power them. In the next section, we describe how physics constrains the data transmission rates from such devices.

### 4.3.1 POWER REQUIREMENTS FOR RECORDING

Any embedded system needs to process data, in preparation for either local storage or wireless transmission. Physics defines hard limits on the required power consumption associated with data processing (neglecting the possibility of reversible logic architectures [205]), arising from the entropy cost for erasing a bit of information [206]:

$$E_{\mathrm{Landauer}} = \ln(2)\, k_{\mathrm{B}} T \approx 3 \times 10^{-21}\, \mathrm{J/bit} \qquad \text{(the Landauer limit)}$$

Ambitious yet physically realistic values for beyond-CMOS logic lie in the tens of $k_{\mathrm{B}}T$ per bit processed [207]. Scaling $40\,k_{\mathrm{B}}T/$bit to record raw voltage waveforms at a minimal $1\,kbit/s/$neuron (e.g. $1\,kHz$ sampling rate, 1 bit processed per neuron per sample), the total power consumption for whole mouse brain recording could in principle be as low as $\sim 16\,nW$. While this leaves $> 10^6$-fold more room (energetically) for increased data processing (more required bit flips per second), or energetic inefficiency of the switching device (greater dissipation per bit), realistic devices in the near-term may in fact require this much overhead, if not more. This necessitates a more detailed consideration of limiting factors for today's microelectronic devices.

In the context of electrical recording, the first step that must be performed by an embedded neural recording device is digitization of the voltage waveform. Until mV-scale switching devices are developed (see discussion below), it is necessary to amplify the $\sim 10$–$100\,\mu V$ spike potential in order to drive digital switching events in downstream gates. During this sub-threshold amplification step, a CMOS (or BJT) device will dissipate static power (associated with a bias current). Importantly, in order to decrease the input-referred voltage noise of this amplification process, it is necessary to increase the bias current and hence the static power dissipation. For a simple differential transistor amplifier, the minimal bias current scales as

$$I_d = \frac{\pi}{2} \frac{4 k_B T}{V_{noise}^2} \frac{k_B T}{q} BW$$

where $V_{noise}$ is the input-referred voltage noise of the amplifier and $q$ is the electron charge. For an extracellular recording with BW = 10 kHz and $V_{noise} = 10 \mu V$, this implies a minimal bias current $I_d \approx 60$ nA or a minimal static power of $(I_d V_{dd}) \approx 6 \times 10^{-8}$ W at $V_{dd} \approx 1$ V operating voltage. Assuming 10 neurons per recording channel, there are then 7.5 million recording channels for a mouse brain, which gives a power dissipation associated with signal amplification of $\sim$500 mW. Note that realistic analog front ends (which are subject to $1/f$ noise and require multiple gain stages) draw 6×–10× greater bias current, quantified by the noise efficiency factor (NEF) [208], to achieve the same input-referred noise levels.

Local on-chip digital computation also incurs an energy cost. Current CMOS digital circuits consume 5–6 orders of magnitude [207, 209–211] more energy per switching event ($\sim$1 fJ/bit including charging of the wires [209]) compared to the Landauer limit (e.g., for a digital CMOS inverter, and ignoring the static power associated with the leakage current). This corresponds to a $\sim$1 fF total load capacitance at 1 V operating voltage. For 100 GHz switching rates ($10^8$ neurons × 1 kHz) as above, this corresponds to 0.01–0.1 mW. Realistic architectures, however, will incur overhead in the number of switching events required to store, compress and/or transmit neural signals, likely bringing the power consumption into an unacceptable range (e.g., 1000 bits processed per sample would be 100 mW here). To take a concrete example, commercial RFID tags consume $\sim$10 $\mu$W [212]. At a chip rate of 256 kbit/s (with a Miller encoding of 2), this yields $7.8 \times 10^{-11}$ J/bit, which is $\sim$10 orders of magnitude higher than the Landauer limit. Applying current RFID technology to whole mouse brain recording at 1 kbit/s/neuron would thus draw $\sim$8 W of power. Therefore, at least 2–3 orders of magnitude reduction in power consumption will be necessary in order to apply embedded electronics for whole-brain neural recording.

Until recently, the energy efficiency of digital computing has scaled on an exponential improvement curve [210]. This was a consequence of Moore's law and Dennard scaling, where both the capacitance of each transistor and its associated interconnect, as well as the operating voltages, were reducing with the device dimensions. Unfortunately, issues related to device variability and the 3D structures needed to maintain the on-to-off current ratio have largely stopped the reduction in effective capacitance per device; current devices are stuck at $\sim$100–200 aF for a minimum sized transistor. Furthermore, the exponential increase in leakage current that comes along with the scaling of the threshold voltage in this scenario has precluded substantial further decreases in voltage at a given performance level. Indeed, for the past several technology generations (since about 2005), CMOS devices have operated at a supply voltage of $\sim$1 V.

While neural signal processing does not demand very stringent transistor speeds and so reductions below $\sim$1 V are certainly feasible, a fundamental limitation in scaling the supply voltage still remains. Specifically, CMOS has a well-defined minimum-energy per bit and an associated minimum-energy operating voltage that is defined by the tradeoff between static (leakage) and dynamic (switching) energy: as the operating voltage is decreased, the capacitive switching energy decreases, but the ratio of currents in the on and off states, $I_{off}/I_{on}$, increases exponentially, increasing the energy associated with leakage (this effect is independent of the threshold voltage in the sub-threshold regime). For practical circuits, the supply voltage that leads to this minimum energy is on the order of 300–500 mV, and thus supply voltage scaling will at most provide 3×–10× improvement in energy over today's designs.

Thus, a paradigm shift in microelectronic hardware is needed to reduce power by several orders of magnitude if we are to approach the physical limits. Developing a switching device operating in the mV range, rather than the 1 V range of current transistors, would allow $(1 V/1 mV)^2 = 10^6$ fold reduction in power consumption [207]. Electronic circuits constructed using analog techniques [213], which sometimes rely on bio-inspired computational architectures, show promise for reducing energy costs by up to five orders of magnitude [213–215], depending on the nature of the computation and the required level of precision.

Figure 4 shows the power consumption per bit processed for several technology classes as well as the corresponding total power consumption required for whole brain readout, assuming a minimal whole-brain bit rate of 100 Gbit/s.

### 4.3.2 POWERING EMBEDDED DEVICES

Embedded systems need power, which could be supplied via electromagnetic or acoustic energy transfer, or could be harvested from the local environment in the brain.

There are two key regimes for wireless electromagnetic power transfer: non-linear device rectification and photovoltaics. If the single-photon energy is sufficient to allow electrons to move from the valence to the conduction band — that is, band gap $< h\nu/q$, where $q$ is the electron charge, $h$ is Planck's constant, and $\nu$ is the frequency of the photon — a photovoltaic effect can occur. Otherwise, electromagnetic energy is converted to voltage by an antenna and non-linear device rectification may occur.

**Figure 4.** Energy cost of elementary operations across a variety of recording and data transmission modalities, expressed in units of the thermal energy (left axis) and as a power assuming 100 GHz switching rate (right axis). The Landauer limit of $k_\mathrm{B}T\ln 2$ sets the minimum energy associated with a logically irreversible bit flip. The practical limit will likely lie in the tens of $k_\mathrm{B}T$ per bit [207], comparable to the free energy release for hydrolysis of a single ATP molecule (or addition of a single nucleotide to DNA or RNA). The energy of a single infrared photon is ∼50 $k_\mathrm{B}T$. Single gates in current CMOS chips dissipate ∼$1 \times 10^5$–$10^6$ $k_\mathrm{B}T$ per switching event, including the capacitive charging of the wires interconnecting the gates (red curve). The switching energy for the gate, not including wires, is ∼100× lower (blue curve). The power efficiency of CMOS has been on an exponential improvement trend due to the miniaturization of components according to Moore's law (data re-digitized from [209]), although power efficiency gains have slowed recently. Current RFID chips compute and communicate at ∼$1 \times 10^9$–$10^{10}$ $k_\mathrm{B}T$ (> 10 pJ) per bit transmitted, while the total energy cost per floating point operation in a 2010 laptop was ∼$1 \times 10^{12}$ $k_\mathrm{B}T$. The power associated with a minimal low-noise CMOS analog front end for signal amplification corresponds to ∼500 mW at whole mouse brain scale. A single two-photon laser pulse at 0.1 nJ pulse energy corresponds to ∼$1 \times 10^{10}$ $k_\mathrm{B}T$. For comparison, the 40 mW approximate maximal allowed power dissipation, according to  above, with its equivalent per-bit energy of ∼$1 \times 10^8$ $k_\mathrm{B}T$ at the minimal 100 Gbit/s bit rate.

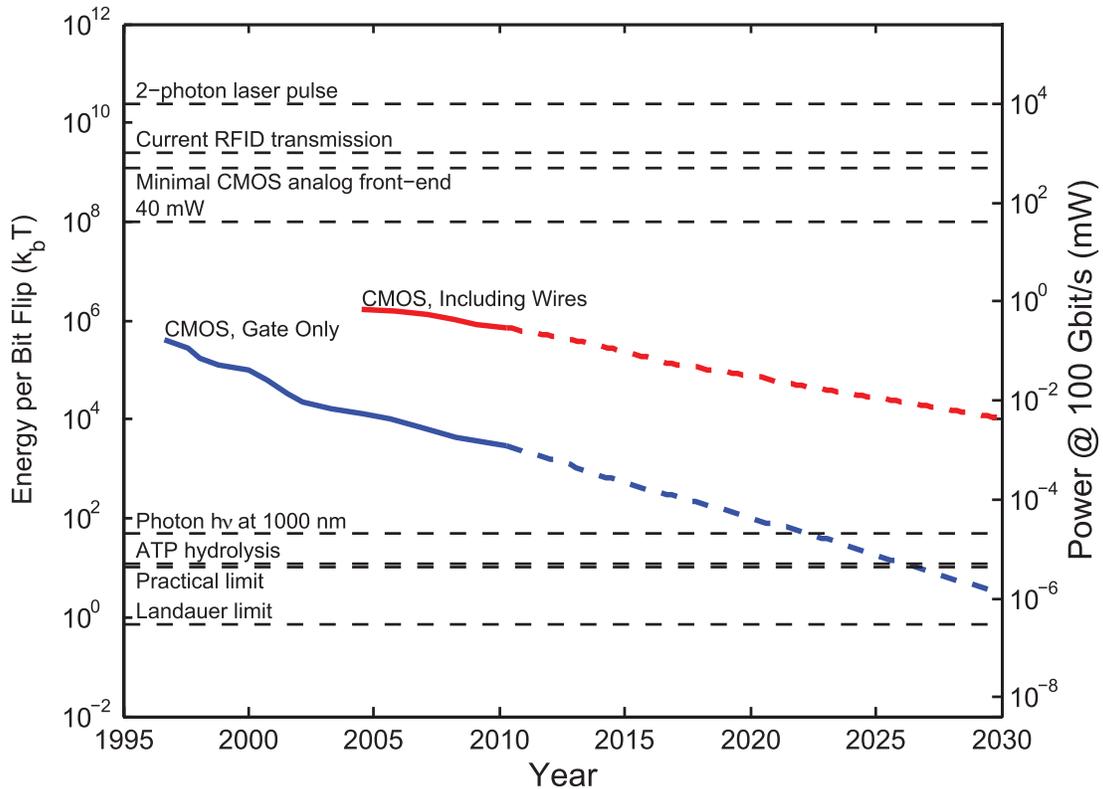

When photon energies are much lower than the band gap, power conversion is governed by the total RF power and by the impedances of the antenna and the rectifier, rather than by the individual photon energy. For a monochromatic RF source, there is no thermodynamic or quantum limit to the RF to DC conversion efficiency, other than the resistive losses and threshold voltages for a semiconductor process. For rectification, when the input voltage to the rectifier is much higher than a semiconductor process threshold, conversion efficiencies of 85 % have been achieved [216]. At low input voltages relative to the semiconductor process threshold, efficiencies as high as 25 % and 2 μW load have been achieved (see [215] for an analysis of power efficiency). Ultimately, rectification improvements are dependent on the same improvements which will be needed for next-generation low-power computing: mV scale switching devices (promising research directions include tunnel FETs [217], electromechanical relays [218] and other options).

While efficient rectification is thus not a fundamental issue, capturing sufficient RF energy in the first place becomes increasingly challenging as microchips become smaller and more deeply embedded in tissue. Wireless electromagnetic power transfer imposes range constraints due to the loss in power density with distance. For directional power transfer, placing the receiver at the edge of the transmitter's near field (the Rayleigh distance $\frac{D^2}{4\lambda}$ where $D$ is the transmitter aperture) has advantages in terms of energy capture efficiency [219], whereas for omni-directional antennas it is advantageous to place the receiver as close as possible to the transmitter. If embedded chips are oriented randomly with respect to the transmitter, the radiation patterns of their antennas cannot be highly directional, i.e., their gains $G_r$ (a measure of directionality) must be close to one. In the far field, this lack of directionality limits power capture by the antenna (due antenna reciprocity [220]): the maximal power $P_A$ available to the chip is

$$P_A = \frac{G_r P_{\text{rad}} \lambda^2}{4\pi}$$

where $P_{\text{rad}}$ is the power density of radiation around the antenna, $\lambda$ is the wavelength and $G_r \approx 1$ for a non-directional antenna [215].

It may be possible to power devices with pure magnetic fields (which are highly penetrant) via near-field (non-radiative) inductive coupling, which is widely used in systems ranging from biomedical implants to electric toothbrushes, or conceivably by using magneto-electric materials [221–224]. For the case of simple inductive coupling, however, the tiny cross-sections of micro-devices limit the amount of power which can be captured: a loop of 10 μm diameter in an applied field of 1 T switching at 1000 Hz produces an induced electromotive force of only 0.1 μV. Assuming a copper loop (~17 nΩ m resistivity) with 1 μm × 1 μm cross-section and 40 μm length (around the outer edge of the chip) gives a power $(V^2/R)$ of only ~15 fW associated with the induced current. In general, the use of coupled high-$Q$ resonators can increase the range and efficiency of near-field electromagnetic power transfer by orders of magnitude [225] compared to non-resonant inductive power transfer and may be particularly relevant for implanted devices [226]. Unfortunately, at the ~10 μm length scale, the achievable on-chip inductances and capacitances are severely limited, which restricts the operating range of any resonant device to high frequencies $\left( f_{\text{resonant}} = \left( 2\pi \sqrt{LC} \right)^{-1} \right)$ which will be attenuated by tissue. Electromagnetic near-field power transfer though tissue to ultra-miniaturized microchips may thus be inefficient, again due to low capture efficiency of the applied fields by tiny device cross-sections.

Alternatively, if the photon energy is above the silicon band gap ($\lambda < \frac{hc}{qV_{\text{th}}} \approx 3\,\mu m$ or less for silicon), the chip is essentially acting as a photovoltaic cell. There is no thermodynamic or quantum limit to the conversion efficiency of light to DC electrical power for monochromatic sources, other than resistive losses and dark currents in the material (86 % in GaAs for example [227]). Again, however, capturing sufficient light becomes difficult for tiny devices. To supply 10 μW (typical of current wirelessly-powered RFID chips) photovoltaically to a 10 μm × 10 μm (cell sized) chip at 34 % photovoltaic efficiency requires a light intensity of ~300 kW/m² at the chip, which is prohibitive. Furthermore, in the use of infrared light for photovoltaics, the penetration of the photons through tissue is decreased compared to radio frequencies.

Piezoelectric harvesting of ultrasound energy by micro-devices is a possibility [70]. The efficiency of electrical harvesting of mechanical strain energy in piezoelectrics can be above 30 % for materials with high electromechanical coupling coefficients (e.g., PZT) [228, 229]. The losses in the piezoelectric transduction process are well described by models such as the KLM model [230, 231].

An alternative to wireless energy transmission is the local harvesting of biochemical energy carriers. Implanted neural recording devices could conceivably be powered by free glucose, the main energy source used by the brain itself. The theoretical maximum thermodynamic efficiency for a fuel cell in aqueous solution is equal to that of the hydrogen fuel cell: $\Delta G^0/\Delta H^0 = 83\,\%$ at 25 °C. Furthermore, if glucose is only oxidized to gluconic acid, the Coulombic (electron extraction) efficiency is at most 8.33 % [232], which bounds the thermodynamic efficiency. The blood glucose concentration in rats has been measured at ~7.6 mM, with an extracellular glucose concentration in the brain of ~2.4 mM [233]. A hypothetical highly miniaturized neural recorder with a device area of 25 μm × 25 μm and efficiency of 80 %, processing a blood flow rate of ~1 mm/s [234] could extract $(80\,\%)(7.6\,\text{mM})(25\,\mu m)^2(1\,\text{mm/s})(2880\,\text{kJ/mol}) \approx 11\,\mu W$, which is sufficient for low-power devices such as RFID chips [235]. Unfortunately, current non-microbial glucose fuel cells obtain only ~180 μW/cm² peak power and ~3.4 μW/cm² steady state power [232]. Thus there is a need for $10^4$- and $10^6$-fold improvements in peak and steady state power densities, respectively, for non-microbial glucose fuel cells to power brain-embedded electronics of the complexity of today's RFID chips (or for the corresponding decrease in power requirements, as emphasized above).

### 4.3.3 Conclusions and Future Directions

The power consumption of today's microelectronic devices is more than six orders of magnitude higher than the physical limit for irreversible computing, and 2–3 orders of magnitude higher than would be permissible for use in whole brain millisecond resolution activity mapping, even under favorable assumptions on the required switching rates and neglecting both the power associated with noise rejection in the analog front end and the CMOS leakage current. Thus, the first priority is to reduce the power consumption associated with embedded electronics. In principle, methods such as infrared light photovoltaics, RF harvesting via diode rectification, or glucose fuel cells, could supply power to embedded neural recorders, but again, significant improvements in the power efficiency of electronics are necessary to enable this. Other potential energy harvesting strategies include materials/enzymes harnessing local biological gradients such as in voltage, osmolarity, or temperature. An analysis of the energy transduction potential of each of these systems is beyond the scope of this discussion. Fortunately, with many orders of magnitude potential for improvement before physical limits are reached, we may expect that embedded nano-electronic devices will emerge as an energetically viable neural interfacing option at some point in the future.

## 4.4 Embedded Devices: Information Theory

Most recording methods envisioned thus far rely on the real-time transmission of neural activity data out of the brain. Physics and information theory impose fundamental limits on this process, including a minimum power consumption required to transmit data through a medium. The most basic of these results hold irrespective of whether the data transmission is wired or wireless, and regardless of the particular physical medium (optical, electrical, acoustic) used as the information carrier.

A communication "channel" is a set of transmitters and receivers that share access to a single physical medium with fixed bandwidth. The bandwidth is the range of frequencies present in the time-varying signals used to transmit information. In wireless communications, information is transmitted by modulating a carrier wave. To allow modulation, the frequency of the carrier wave must be higher than the bandwidth: for example, a 400 THz visible light wave may be modulated at a 100 GHz rate. The physical medium underlying a channel could be a wire (with a bandwidth set by its capacitive RC time constant), an optical fiber, free space electromagnetic waves over a certain frequency range, or other media.

As a concrete example, consider a police department with 100 officers, each possessing a hand-held radio. The radios transmit vocalizations by modulating an 80 MHz carrier wave at ∼10 kHz. This constitutes a single shared communications channel with 10 kHz bandwidth. Simultaneously, the fire department may communicate via a separate channel, also with a bandwidth of ∼10 kHz, by modulating a 90 MHz carrier wave. The channels are separate because modulation introduced into one does not affect the other. If the neighboring town's police department makes the mistake of also operating at 80 MHz carrier frequency, then they share a channel and conflicts will arise.

### 4.4.1 Power Requirements for Single-Channel Data Transmission

We first treat the case in which there is a single channel for transmitting data out of the brain. As discussed above in the context of electrical spike sorting, the Shannon Capacity Theorem [90] sets the maximal bit rate for a channel (assuming additive white Gaussian noise) to

$$R_{max} = BW \log_2 \left(1 + SNR\right)$$

where BW is the channel bandwidth and SNR is the signal-to-noise ratio. If there is only thermal noise the $SNR = P/(N_0 BW)$, where $N_0$ is the thermal noise power spectral density of $k_B T$ W/Hz and $P = (PL)P_0$ is the power of the transmitted signals $P_0$, weakened by path loss PL. Therefore the transmitted power $P_0$ is lower-bounded:

$$P_0 > k_B T \; BW \; \frac{2^{R_{max}/BW} - 1}{PL}$$

as shown in Figure 5 (bottom). In a minimal model of a transmitter-receiver system, there thus exists a tradeoff between the required signal power and the bandwidth of the carrier radiation, due to the thermal noise floor, even in the absence of path loss (PL = 1).

Path loss weakens the proportion of the power that can reach the detector. Using the above equation, we can calculate, as a function of bandwidth, the power necessary to transmit a target whole-brain bit rate of 100 Gbit/s through a medium with path loss dependent on the carrier wavelength, as shown in Figure 5 (top).

For RF wavelengths, the radiation penetrates deeply but the achievable data rates are low without excessive power consumption, due to the limited bandwidth. For wavelengths intermediate between RF and infrared, the penetration depth is low and power must be expended to combat these losses, despite the high carrier bandwidth. Only in the infrared and visible ranges do the tradeoffs between power, bandwidth and penetration depth allow transmission of > 100 Gbit/s out of the brain through a single channel without unacceptable power consumption.

**Figure 5.** Power requirements imposed by information theory on data transmission through a single (additive white Gaussian noise) channel with carrier frequency *v* (an upper bound on the bandwidth), given thermal noise and path loss. Bottom: absorption length of water as a function of frequency (blue), minimal power to transmit data at 100, 1000 and 10 000 Gbit/s (green) as a function of frequency, assuming thermal noise but no path loss. Top: minimal power to transmit data at 100, 1000 and 10 000 Gbit/s as a function of frequency, assuming thermal noise and a path loss corresponding to the attenuation by water absorption over a distance of 2 mm. While formulated for a single channel, at certain wavelengths (e.g., RF) these factors also constrain multiplexed data transmissions between many transmitters and many receivers, depending on capacity of the system for spatial multiplexing. Horizontal dashed lines: 40 mW, the approximate maximal whole-brain power dissipation in steady state.

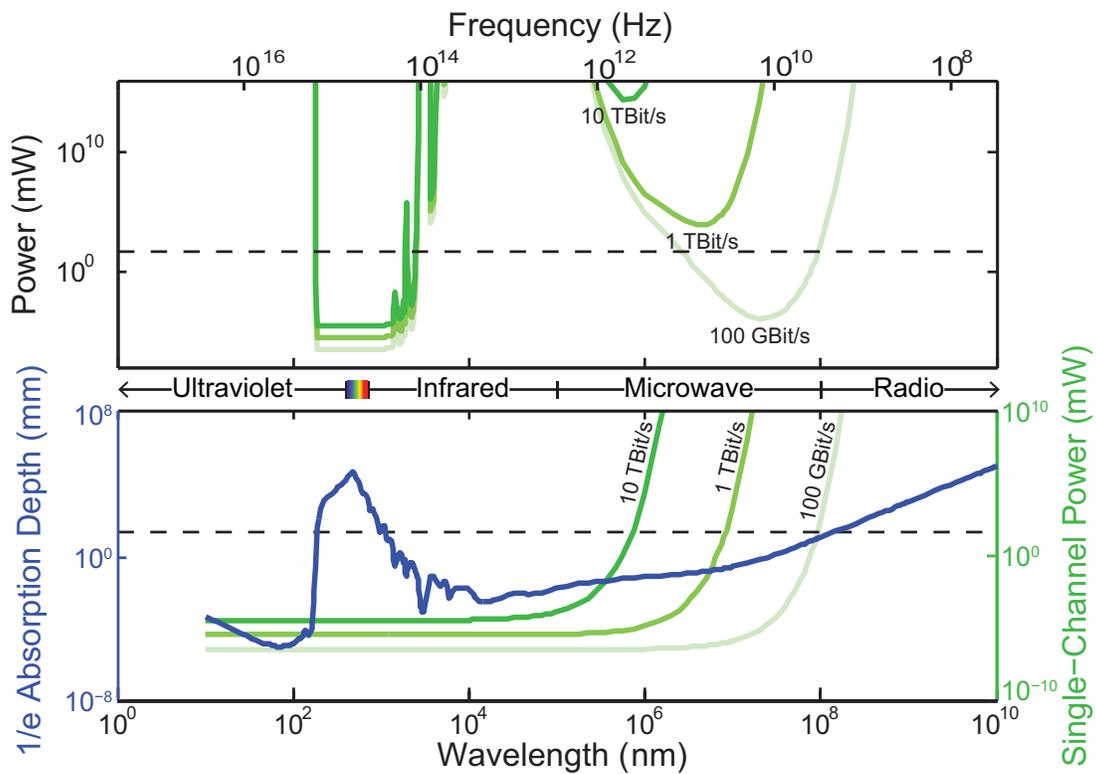

The analysis above has ignored the effects of noise sources other than thermal noise, but many additional noise sources will increase the amount of power needed to transmit data, via a decrease in the SNR at fixed input power. For optical transmission in the brain, the noise is dominated by time-correlated "speckle noise" below 200 kHz, which arises mostly from local blood flow [236]. This correlated noise, which cannot be filtered by simple averaging, could be avoided by modulating optical signals at frequencies above 200 kHz.

### 4.4.2 SPATIALLY MULTIPLEXED DATA TRANSMISSION

As discussed above, transmitting information through a single channel imposes direct limits on bit rate, carrier frequency and input power. However, it is conceivable to divide the data transmission burden over many independent channels, i.e., over many pairs of transmitters and receivers, each operating at lower bandwidth (e.g., at radio frequencies). Indeed, this would be optimal in a scenario where many embedded devices measure and then transmit the activities of nearby neurons. As a concrete example of such "spatial multiplexing," an effective capacity of 1 Tbit/s could conceivably be obtained by splitting the data over 1000 transmitter-receiver pairs each operating at 1 Gbit/s, with the transmitters arranged in a 10 × 10 × 10 grid. Importantly, in order to exceed the above limits for single-channel data transmission, it must be possible for these transmitter receiver pairs to share the same bandwidth and operate simultaneously without conflicts, for example by modulating distinguishable carrier waves or by transferring data over separate wires. The conditions under which this may occur, however, can be counter-intuitive. For example, for antennas to operate independently, they must be spaced apart from one another by roughly a wavelength. For 10 GHz microwaves, the wavelength is ~3 cm, so no more than a handful of microwave transmitters (e.g., operating at frequencies in the 100 GHz–1 THz range) can co-occupy the mouse brain while operating independently.

Even with many non-independent transmitters co-occupying the brain and operating simultaneously over the same frequency spectrum, it may be possible under some conditions to "factor out" the effects of the coupling and allow an increase in channel capacity relative the single-channel result. To treat such scenarios, a generalization to Shannon's capacity theorem to multi-input-multi-output (MIMO) channels has shown that the maximal total data rate is

$$R_{\max} = \text{BW} \cdot \log_2 \left| \boldsymbol{I} + (\text{SNR}) \boldsymbol{H} \boldsymbol{H}^* \right|$$

where $\boldsymbol{I}$ is the identity matrix, $|\cdot|$ denotes the matrix determinant, $\boldsymbol{H}$ is the ($M \times N$ for $N$ transmitters and $M$ receivers) channel matrix giving the coupling between the vector of transmitted signals and the vector of received signals and $\boldsymbol{H}^*$ denotes the matrix adjoint of $\boldsymbol{H}$ [237]. The vector of received signals is then $\boldsymbol{y} = \boldsymbol{H}\boldsymbol{x} + \boldsymbol{n}$ where $\boldsymbol{x}$ is the vector of transmitted signals and $\boldsymbol{n}$ is a noise vector. Any matrix can be written as $\boldsymbol{H} = \boldsymbol{U}\boldsymbol{\Sigma}\boldsymbol{V}^*$ where $\boldsymbol{U}$ and $\boldsymbol{V}$ are unitary matrices, and $\boldsymbol{\Sigma}$ is a diagonal matrix whose elements are the *singular values* $\lambda_i$. One can re-write the above equation as

$$R_{\max} = \text{BW} \cdot \sum_{i=1}^{\min(M,N)} \log_2 \left( 1 + \text{SNR} \cdot \lambda_i^2 \right)$$

If the matrix $\boldsymbol{H}$ is of full rank, then the capacity for the multi-channel system can increase over the single-input-single-output (SISO) result by $\min(M,N)$ times [238]. Note that the rank of the matrix corresponds to the number of non-zero singular values, so an analysis of the singular values of channel matrices can inform us about the multiplexing capacity of the channel. Furthermore, this multiplexing capacity can in principle be achieved even when the transmitters are not in communication with each other, which could potentially be important for scenarios involving many brain embedded transmitters [239].

Transmission through a medium with negligible scattering is the simplest situation to analyze. In this case, evaluating the matrix $\boldsymbol{H}$ requires knowledge of the transmitter-transmitter, transmitter- receiver, and receiver-receiver distances, as well as the orientations and radiation patterns of the antennas (e.g., high gain antennas will have a highly directional radiation pattern). Depending on these factors, the beam from each transmitter will spread to impinge upon multiple receivers and the effective number of spatially independent beams will be reduced. With transmitter-transmitter and receiver-receiver distances larger than the wavelength, and highly directional antennas with appropriately chosen orientations, it is possible to increase the channel capacity linearly with $\min(M,N)$.

Random scattering, in a coherent disordered medium where the mean free-path $\ell$ is much larger than the wavelength $\lambda$ and much smaller than the size of the disordered medium, is another condition where the matrix $\boldsymbol{H}$ is a random scattering matrix of full rank [240, 241]. Intuitively, for the case of two transmitters and two receivers separated by a disordered medium larger than the mean free path: if transmitter 1 is at least a mean-free path from transmitter 2 (or potentially as close as a few wavelengths [242]), the path from transmitter 1 to receiver 1 and the path from transmitter 2 to receiver 2 would be uncorrelated with respect to one another (in terms of physical path, phase, amplitude fluctuations, and other properties). The rank of the matrix $\boldsymbol{H}$ would then be 2. Devising a code on the transmitter such that the receivers can distinguish between these two uncorrelated streams results in a doubling of the capacity, rather than simply averaging the noise floor, which would provide only a logarithmic capacity gain due to the increased SNR.

Thus, contrary to intuition, a high degree of random scattering can potentially be useful for data transmission, by enabling spatial multiplexing of channels. This idea has been demonstrated experimentally in the context of ultrasound transmissions [243]. Biological tissue in the infrared range is well described as such a random scattering medium (e.g., mean free path $\sim 200\,\mu m$ at $\sim 800\,nm$ *in vivo*). Therefore infrared light could be used for spatially multiplexed data transmission out of the brain. At wavelengths $\lambda$ comparable to critical brain dimensions in the mouse, however, an insufficient number of scattering events will occur to create multiple independent pathways for $N$ transmitters. Mathematically, the matrix $H$ will have one highly dominant singular value and a number of much smaller remaining terms, such that the signals appearing at a receiver from two separate transmitters will be highly linearly dependent, differing only by a small phase angle. Therefore, there will be no capacity gain from multiple transmitters, and distinct transmitters will effectively share a single channel (reducing to the SISO result).

Little is known about the biological interaction with electromagnetic fields at wavelengths much shorter than the critical brain dimensions but beyond the infrared, approximately 100 GHz ($\sim 3\,mm$) to 100 THz ($\sim 3\,\mu m$) in the mouse. If multiple scattering occurs and the absorption is low, this may also be a regime conducive to MIMO communications [244]. Efficiently generating and processing radiation in this regime by embedded devices is an outstanding problem, however. The so-called "THz-gap" [245] exists because (moving towards higher frequencies starting from DC electronics), parasitic capacitances and passive losses limit the maximum frequency at which a field-effect transistor (FET) may oscillate and on the other hand (moving downward in frequency starting from optics), the band-gaps of opto-electronic devices limit the minimum frequency at which quantum transitions occur. Thus there is no high-power, low-cost, portable, room temperature THz source available. Advances in THz light generation, e.g. through the use of tunneling transistors, could be enabling.

### 4.4.3  Ultrasound as a Data Transmission Modality

An important caveat to these conclusions on wireless data transfer occurs if we consider the use of ultrasound rather than electromagnetic radiation. Because the speed of sound is dramatically slower than that of light, the wavelength of 10 MHz ultrasound is only $\sim 150\,\mu m$ (approximating the speed of sound in brain as the speed of sound in water, $\sim 1500\,m/s$). Thus, many 10 MHz ultrasound transmitters/receiver could be placed inside a mouse brain while maintaining their spatial separation above the wavelength, and a linear scaling of the MIMO channel capacity with the number of devices is likely possible in this regime, assuming that appropriate antenna gains and orientations can be achieved inside brain tissue. Beam orientation could present a challenge if micro-devices are oriented randomly after implantation. With an attenuation of 0.5 dB/(cm MHz) [246], the attenuation at 10 MHz is only 5 dB/cm. Thus ultrasound-based transmission of power and data from embedded recording devices may be viable [70].

In contrast, direct imaging of neural activity by ultrasound (e.g., using contrast agents which create local variations in tissue elastic modulus or density) may be more difficult. While the theoretical (diffraction-limited) and currently practical resolutions of 100 MHz ultrasound are $\sim 15\,\mu m$, and 15–60 $\mu m$ [247], respectfully, at these frequencies, power is attenuated by brain tissue with a coefficient of $\sim 50$ dB/cm [246] ($10^5$-fold attenuation per cm), which imposes a penetration limit (e.g., for measurements with a dynamic range of 80 dB [247]). Attenuation of ultrasound by bone is stronger still, at 22 dB/(cm MHz) [246]. Attenuation could therefore limit the use of ultrasound as a high-resolution neural recording modality in direct imaging modes, but multiplexed transmission of lower-frequency ultrasound from embedded devices could sidestep this issue.

### 4.4.4  Conclusions and Future Directions

Physics and information theory impose a tradeoff between bandwidth and power consumption in sending data through any communication channel. Considering only thermal noise and no path loss, achieving 100 Gbit/s data rates through a single channel necessitates either a bandwidth above a few GHz or a transmitted power above $\sim 100$ mW, the latter of which may be prohibitive from a heat dissipation perspective if the signals are to be generated by dissipative microelectronic devices. Researchers have proposed to use thousands or millions of tiny [248] wireless transmitters embedded in the brain to transmit local neural activity measurements to an external receiver via microwave radiation [249]. However, based on the above power-bandwidth tradeoff, this will require a bandwidth above a few GHz. At the corresponding carrier frequencies, the penetration depth of the microwave radiation drops significantly, requiring increased power to combat the resulting signal loss. While one might hope that multiple independent channels could be multiplexed inside the brain, reducing the bandwidth and power requirements for each individual channel, the long wavelengths of microwave radiation compared to the mouse brain diameter suggest that such channels cannot be independent, as is confirmed by an analysis of the multi-input-multi-output (MIMO) channel capacity for this scenario. Therefore, radio-frequency electromagnetic transmission of whole brain activity data from embedded devices does not appear to be a viable option for brain activity mapping.

On the other hand, an analysis of the channel capacity for IR transmissions in a diffusive medium suggests that, because of its high frequency and decent penetration depth, infrared radiation may provide a viable substrate for transmitting activity data from embedded devices. For example, data could be transmitted via modulating the multiple-scattering speckle pattern of infrared light by varying the backscatter from an embedded optical device, such as an LCD pixel [250], in an activity-dependent fashion. Because the speckle pattern is sensitive to the motion of a single scatterer [242, 251], coherent multiple scattering could effectively act as an optical amplifier and as a means to create independent communication pathways. Furthermore, multiplexed data transmission via ultrasound is likely possible because of its short wavelength in tissue at reasonable carrier frequencies. It may also be of interest to explore network architectures [252] in which data is transmitted at low transmit power over short distances via local hops between neighboring nodes capable of signal restoration.

## 4.5 MAGNETIC RESONANCE IMAGING

Magnetic resonance imaging (MRI) uses the resonant behavior of nuclear spins in a magnetic field to non-invasively probe the spatiotemporally varying chemical and magnetic properties of tissues. Although originally conceived as a means to image anatomy, MRI can be used to observe neural activity provided that correlates of such activity are reflected in dynamic changes in local chemistry or magnetism.

In an MRI study, a strong static field ($B = 1$–$15\,\mathrm{T}$) is applied to polarize nuclear spins (usually $^1\mathrm{H}$), causing them to resonate at a field-dependent Larmor frequency

$$f = \frac{\gamma}{2\pi} B$$

where $\gamma$ is the gyromagnetic ratio of the nucleus (e.g., $^1\mathrm{H}$ has a gyromagnetic ratio of $267.522\,\mathrm{MHz/T}$ [253] and therefore resonates at $42.577\,\mathrm{MHz}$ in a $1\,\mathrm{T}$ field). To obtain positional information, spatial field gradients are applied such that nuclei at different positions in the sample resonate at slightly different frequencies. Sequences of RF pulses and gradients are then applied to the sample, eliciting resonant emissions that contain information about spins' local chemical environment, magnetic field anisotropy and various other properties.

Most functional studies rely on dynamic changes in two forms of relaxation experienced by RF-excited spins. The first form results from energy dissipation through interactions with other species (e.g. other spins or unpaired electrons), causing the spins to recover their lowest energy state on a timescale, $T_1$, of $100$–$1000\,\mathrm{ms}$ [254]. The second form of relaxation reflects the dephasing of spin signals in a given sampling volume (voxel) over a timescale, $T_2$, of $10$–$100\,\mathrm{ms}$ [255] due to non-uniform Larmor frequencies caused, e.g., by the presence of local magnetic field inhomogeneities.

In blood-oxygen level dependent [256] functional MRI (BOLD-fMRI), the most widely used form of neural MR imaging, increased neural activity in a given brain region alters the vascular concentration of paramagnetic deoxy-hemoglobin, which affects local magnetic field homogeneity and thereby alters $T_2$. Although the existence of this paramagnetic reporter of oxygen metabolism is fortuitous, the data it provides is only an indirect readout of neural activity [257–259], which is limited in its spatial and temporal resolution to the dynamics of blood flow in the brain's capillary network (1–2 s). The spatial point-spread function of the hemodynamic BOLD response is in the 1 mm range, although sub-millimeter measurements, revealing cortical laminar and columnar features, have been obtained by filtering out the signals from larger blood vessels [260]. A significant area of current and future work is aimed at developing new molecular reporters that can be introduced into the brain to transduce aspects of neural signaling such as calcium spikes and neurotransmitter release into MRI- detectable magnetic or chemical signals [261–263], as described in section 4.5.3, below.

### 4.5.1 SPATIOTEMPORAL RESOLUTION

The temporal resolution of MRI is limited by the dynamics of spin relaxation. For sequential MR signal acquisitions to be fully independent, spins must be allowed to recover their equilibrium magnetization on the timescale of $T_1$ (100–1000 ms). However, if local $T_1$ is static its pre-mapping could enable temporally variant $T_2$ effects to be observed at refresh rates on the faster $T_2$ timescale (10–100 ms) [255]. It may also be possible to detect events that occur on a timescale shorter than $T_1$ and $T_2$, if the magnitude of the resulting change in spin dynamics overcomes the lack of independence between acquisitions. Note that these limitations on the repetition time of the underlying pulse sequence are not eliminated by "fast" pulse sequences such as echo-planar imaging (EPI) [264] and fast low-angle shot (FLASH) [265] or by the use of multiple detector coils [266]. These techniques accelerate the acquisition of 2D and 3D images, but still require spins to be prepared for readout.

The spatial resolution of current MRI techniques is limited by the diffusion of water molecules during the acquisition time [267], since contrast at scales above the diffusion length will be attenuated by diffusion. The RMS distance of a water molecule from its origin, after diffusing in 3D for a time $T_{\mathrm{acq}}$, is

$$d_{\mathrm{rms}} = \sqrt{6 D_{\mathrm{water}} T_{\mathrm{acq}}}$$

where $D_{\mathrm{water}} = 2300\,\mu\mathrm{m}^2/\mathrm{s}$ is the self-diffusion coefficient of water. For $T_{\mathrm{acq}} \approx 100\,\mathrm{ms}$, $d_{\mathrm{rms}} \approx 37\,\mu\mathrm{m}$, which sets the approximate spatial resolution. For ultra-short acquisitions at $T_{\mathrm{acq}} \approx 10\,\mathrm{ms}$, $d_{\mathrm{rms}} \approx 12\,\mu\mathrm{m}$.

More technically, as described above, MRI uses field gradients to encode spatial positions in the RF frequency (wavenumber) components of the emitted radiation. The quality of the reconstruction of frequency space thus limits the achievable spatial resolution. The sampling interval of the detector $\Delta t$, and the field gradient $G$, determine the wavenumber increment as

$$\Delta k = \gamma G \Delta t$$

The spatial resolution (here considering only one dimension) is then given by [267]:

$$\Delta x_{k\text{-space}} = \frac{\pi}{\frac{T_{\text{acq}}}{\Delta t} \Delta k} = \frac{\pi}{T_{\text{acq}} \gamma G}$$

Note that it is the gradient field, not the polarizing field $B_0$, which determines the resolution. For a gradient field of 100 mT/m and an acquisition time of 100 ms

$$\Delta x_{k\text{-space}} = \frac{\pi}{(100\,\text{ms})\,(267\,\text{MHz/T})\,(100\,\text{mT/m})} \approx 1.17\,\mu\text{m}$$

Due to relaxation, however, the emissions from a spin at a given position do not constitute a pure tone with a well-defined frequency. Instead, each spin exhibits a frequency spread, which gives rise to another limit on the spatial resolution [267]:

$$\Delta x_{\text{relaxation}} = \frac{2}{\gamma G T_2^*}$$

where $T_2^*$ is the shortest relaxation time. Assuming $T_2^* = 5\,\text{ms}$ and $G = 100\,\text{mT/m}$, gives

$$\Delta x_{\text{relaxation}} \approx 14\,\mu\text{m}$$

Therefore, for water protons, the resolution limit is set by diffusion over $\sim 100\,\text{ms}$ acquisition timescales, rather than by k-space sampling or relaxation. For other spin species (e.g., with lower diffusion rate), it may be possible to achieve resolutions limited by frequency discrimination.

Notably, there exists a practical trade-off between spatial resolution, temporal resolution, and sensitivity (SNR). In particular, to achieve high spatial resolution, it is necessary to densely sample $k$-space. Fast sampling sequences such as FLASH and EPI achieve speed by sampling each point of $k$-space using less signal and often at a lower resolution. Even at high field strengths (11.7 T), this tradeoff results in practical EPI-fMRI with a spatial resolution of $150\,\mu\text{m} \times 150\,\mu\text{m} \times 500\,\mu\text{m}$ and a temporal resolution of 200 ms [268]. Achieving much higher spatial resolutions requires longer acquisitions and/or lower temporal sampling. For example, achieving a $20\,\mu\text{m}$ anatomical resolution in MRI of *Drosophila* embryos required 54 minutes for a small field of view of $2.5\,\text{mm} \times 2.5\,\text{mm} \times 5\,\text{mm}$ [269]. Furthermore, the flies were administered paramagnetic gadolinium chelates to shorten $T_1$ and thereby the acquisition time. Separately, frame rates of 50 ms have been obtained for dynamic imaging of the human heart, but required the use of strong priors to reduce data collection requirements [270].

### 4.5.2 Energy Dissipation

Energy is dissipated into the brain when the excited spins relax to their equilibrium magnetization in the applied field. The energy associated with this relaxation is of order the Zeeman energy:

$$\Delta E_{\text{Zeeman}} = \frac{\gamma}{2\pi} h B_0$$

To obtain an upper bound on the heat dissipation of MRI, we first assume that the brain is entirely water, that every proton spin is initially aligned by the field and then excited by the RF pulse, and that all spins relax during a $T_1$ relaxation time of $\sim 600\,\text{ms}$. In this scenario, even an applied field of as high as $\sim 200\,\text{T}$ would generate dissipation within the $\sim 50\,\text{mW}$ energy dissipation limit. In reality, the energy dissipation is 4–5 orders of magnitude smaller, because only a tiny fractional excess of the spins are initially aligned by the field ($\sim 1 \times 10^{-5}$ for fields on the order of 1 T). Therefore, thermal dissipation associated with spin excitation in MRI is unlikely to cause problems unless field strengths much greater than the largest currently used fields ($\sim 20\,\text{T}$) are invoked, or spins with much higher gyromagnetic ratios are used.

Practically, the main energy consideration in MRI is the absorption by tissues of RF energy applied during imaging pulse sequences and the switching of magnetic field gradients. Such absorption is often calculated through numerical solutions of the Maxwell Equations taking into account the precise geometry, tissue properties and applied fields for a particular experimental setup [271]. The typical specific absorption rate (SAR) is well under 10 W/kg (or 5 mW per 500 mg), and is restricted by the FDA to less than 3 W/kg for human studies.

### 4.5.3 IMAGING AGENTS

All the preceding discussion about spatiotemporal resolution presumes the existence of local time-varying signals (e.g., changes in $T_1$ or $T_2$) corresponding to the dynamics of neural activity. The hemodynamic BOLD response is the most prominent such signal, the limitations of which are discussed above. There have been studies working towards direct detection of minute (e.g., $\sim 0.2\,\text{nT}$) magnetic fields associated with action potentials through their effects on MRI phase or magnitude contrast [272, 273], but reliably detecting these fields above the physiological noise will likely require novel strategies [274, 275] and estimates of the feasibility of these methods have been complicated by the lack of a realistic model for the local distribution of neuronal currents. MRI detection of the mechanical displacement of active neurons due to the Lorentz force in an applied magnetic field [276] has also been explored, as has the detection of activity-dependent changes in the diffusion of tissue water [277, 278], possibly due to neuronal or glial [279] cell swelling [280, 281], although strongly diffusion-weighted scans may have disadvantages in terms of SNR [282]. Manganese influx through voltage-gated calcium channels [283, 284] generates MRI contrast, but exhibits slow uptake kinetics and even slower efflux, such that manganese monotonically accumulates in the neurons over time. Conceivably, over-expression of manganese efflux pumps such as the iron transporter ferroportin [285] could allow time-dependent activity imaging using manganese contrast.

In the past 15 years, efforts have been undertaken to develop chemical and biomolecular imaging agents that can be introduced into the brain to produce MRI detectable signals corresponding to specific aspects of neural function (analogously to fluorescent dyes and proteins). One critical advantage of using genetically encoded indicators would be the ability to target these indicators to specific cell types [286, 287] and/or cellular compartments [186–193]. Notable examples of engineered molecular MRI contrast agents include $T_1$ and $T_2$ sensors of calcium [288, 289] and a $T_1$ sensor of neurotransmitter release [261]. Depending on their mode of action, these imaging agents can provide temporal resolutions ranging from 10 ms to 10 s [290]. However, a major current limitation for fast agents is the requirement that they be present in tissues at $\mu$M concentrations, posing major challenges for delivery and genetic expression. Model organisms lacking hemoglobin (e.g., the blowfly), and hence lacking a hemodynamic BOLD response (as is also the case for ex-vivo brain slices), may be particularly useful for in-vivo testing of novel activity-dependent contrast mechanisms, and specialized setups have been constructed to perform MRI at near-cellular spatial resolution in this context (though still requiring several hours to generate whole-brain anatomical images at this resolution) [291].

Figure 6 shows the achievable temporal resolution for various classes of activity-dependent MRI contrast agents as well as the spatial resolution limit due to water proton diffusion.

**Figure 6.** Key factors determining the spatiotemporal resolution of dynamic MRI imaging. (a) Temporal resolution and contrast agent concentration allowing > 5 % contrast, for different classes of dynamic MRI contrast agent (reproduced from [290], with permission). (b) Diffusion limited spatial resolution for water proton MRI as a function of temporal resolution.

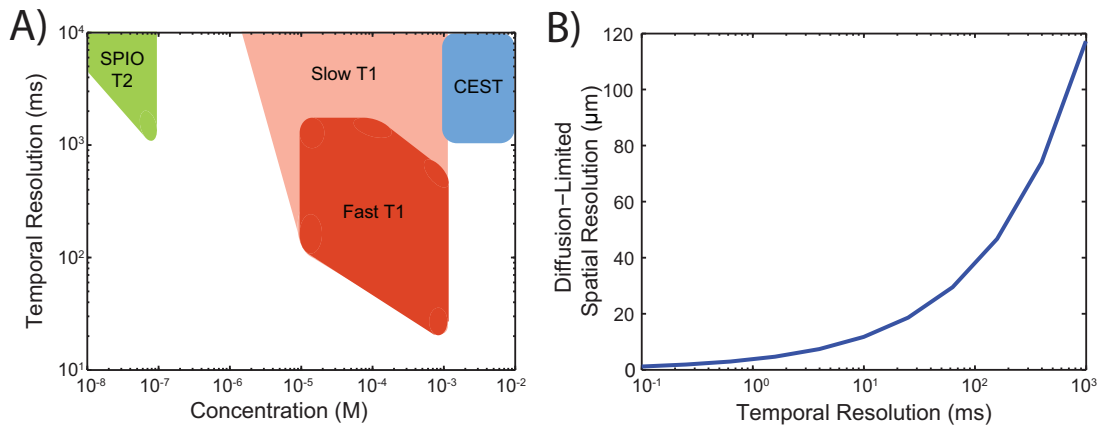

### 4.5.4 CONCLUSIONS AND FUTURE DIRECTIONS

Moving beyond hemodynamic contrast is crucial for improving the spatiotemporal resolution of fMRI, and several avenues may be available for doing so, especially through the use of novel molecular contrast agents and/or genetic engineering. More fundamentally, current MRI techniques rely on the excitation of proton spins in water: this limits imaging to > 100 ms timescales, unless SNR is severely compromised, due to the low polarizability and long $T_1$ relaxation times of proton spins. There is also a spatial resolution limit of tens of microns over these timescales due to water's fast diffusion. Methods which couple neural activity to non-diffusible, highly polarized spins could, in principle, ameliorate this situation.

## 4.6 Molecular Recording

An alternative to electrical, optical or MRI recording is the local storage of data in molecular substrates. Each neuron could be engineered to write a record of its own time-varying electrical activities onto a biological macromolecule, allowing off-line extraction of data after the experiment. Such systems could, in principle, be genetically encoded, and would thus naturally record from all neurons at the same time.

One proposed implementation of such a "molecular ticker tape" would utilize an engineered DNA polymerase with a $Ca^{2+}$-sensitive or membrane-voltage-sensitive error-rate [5] to record time-varying neural activities onto DNA [6] as patterns of nucleotide misincorporations relative to a known template DNA strand (for alternative local recording techniques see [292, 293]). The time-varying signal would later be recovered by DNA sequencing and subsequent statistical analysis [6]. DNA polymerases found in nature can add up to ~1000 nucleotides per second [294], and certain non-replicative polymerases such as DNA polymerase iota have error rates of > 70 % on template T bases [295]. Similar strategies could be implemented using RNA polymerases or potentially using other enzyme/hetero-polymer systems.

### 4.6.1 Spatiotemporal Resolution

Polymerases proceed along their template DNA strands in a stochastic, thermally driven fashion; thus, polymerases that are initially synchronized will de-phase with respect to one another over time, occupying a range of positions on their respective templates at the time when a neural impulse occurs. The rate of this de-phasing is a key parameter governing the temporal resolution of molecular recording. By averaging over many simultaneously replicated templates, it is theoretically possible to associate variations in nucleotide misincorporation rate with the times at which these variations occurred, and thus to obtain temporally resolved recordings of the cation concentration [6].

An analysis of the projected temporal resolution of molecular ticker tapes as a function of polymerase biochemical parameters can be found in [6]. This work suggests that molecular ticker tapes require synchronization mechanisms if they are to record at < 10 ms temporal resolution for durations longer than seconds, even when 10000 templates per cell are recorded simultaneously, unless engineered polymerases with kinetic parameters beyond the limits of those found in nature can be developed. Recording at lower temporal resolutions, however, appears feasible using naturalistic biochemical parameters, even in the absence of synchronization mechanisms.

The development of mechanisms to improve synchronization of the ensemble of polymerases within each cell, or to encode timestamps into the synthesized DNA (e.g., molecular clocks), could improve temporal resolution and decrease the number of required template strands per neuron. Mutation-based molecular clocks over evolutionary timescales are widely used in the field of phylogenetics [296], and new tools from synthetic biology [297] and optogenetics or thermogenetics [298] also suggest strategies for building molecular clocks on faster timescales. As an example sketch of a possible synchronization mechanism, optogenetic methods (e.g., similar to [299]) could be used to halt, and thus re-phase, a sub-population of polymerases at a light-dependent pause site in the template DNA, while another sub-population of polymerases reads through this pause site to maintain temporal continuity of recording; then the second population could be re-synchronized at an orthogonal light-dependent pause site while the first population reads through. Alternatively, some form of optogenetics could be used to directly write bit strings encoding time stamps into the synthesized DNA. These strategies would require one or two, sufficiently strong global clock signals to be optically broadcast to all neurons. The optics involved would be comparatively simple: this could be done using far fewer optical fibers than would be required for fiber-based activity readout, for instance. Alternatively, if the brain could be flash-frozen at a precisely known time, this could serve as a global time-stamp corresponding to the termination of DNA synthesis (e.g., the DNA 3' end).

Spatial resolution for molecular recording would naturally reach the single cell level. To determine which nucleic acid tape originated from which neuron, static cell-specific DNA barcoding could be used [300] to associate the synthesized DNA strands with nodes in a topological connectome map obtained via DNA sequencing. Fluorescence in-situ DNA sequencing (FISSEQ) [301] on serially-sectioned or intact tissue (fixed post-mortem) [302] could be used to obtain explicit geometric information.

### 4.6.2 Energy Dissipation

**Nucleotide metabolism** DNA polymerization imposes a metabolic load on the cell. Replication of the 3 billion bp human genome takes approximately eight hours in normally dividing cells, which equates to a nucleotide incorporation rate of ~100 kHz. Therefore, in order not to exceed the metabolic rates associated with normal genome replication, molecular ticker tapes operating at 1 kHz polymerization speed [294] would be limited to approximately 100 simultaneously replicated templates per cell. Even more recordings would be possible for RNA ticker tapes. The mammalian cell polymerizes at least $10^{11}$ NTPs per 16-hour cell cycle [303]. Therefore, ~1700 RNA tickertapes, each operating at 1 kHz, could be placed in a cell before generating a metabolic impact equal to that of the cell's baseline transcription rate. While these comparisons to baseline physiological levels are reasonable guidelines, it is likely that a neuron can support higher metabolic loads associated with larger numbers of templates. The maximal rate of neuronal aerobic respiration is ∼5 fmol of ATP minute via oxidative respiration (see the section on bio-luminescence). Assuming ∼1 ATP equivalent consumed per nucleotide incorporation, if neuronal metabolism were entirely dedicated to polymerization, it could support the incorporation of up to $6 \times 10^9$ nucleotides per minute, or $10^5$ simultaneously replicated DNA templates at 1 kHz.

**Power dissipation**   Normal DNA and RNA synthesis do not produce problematic energy dissipation and molecular tickertapes will likewise not be highly dissipative, at least in the regime where nucleic acid polymerization rates do not exceed those associated with genome replication or transcription.

### 4.6.3   Volume Displacement

The nucleus of a neuron occupies ∼6 % of a neuron's volume $((4\,\mu m)^3/(10\,\mu m)^3)$. Ticker tapes operating at 1 kHz with 10 000 simultaneously replicated templates could record for 300 seconds before the total length of DNA synthesized equals the human genome length. In the case of RNA polymerase II-based transcription, 2.75 h of recording by 10 000 recorders is required to reach the net transcript length in the cell. Therefore, with appropriate mechanisms to fold/pack the nucleic acids generated by molecular ticker tapes, they would not impose unreasonable requirements on cellular volume displacement over minutes to hours.

### 4.6.4   Conclusions and Future Directions

Molecular recording of neural activity has the advantages of inherent scalability, single-cell precision, and low energy and volume footprints. Making molecular recording work at temporal resolutions approaching 1 kHz, however, will require multiple new developments in synthetic biology, including protein engineering to create a fast polymerase (> 1 kHz) that strongly couples proxies for neural activity to nucleotide incorporation probabilities. Synchronization mechanisms would likely be required to perform molecular recording at single-spike temporal resolution. An attractive potential payoff for molecular approaches to activity mapping is the prospect of seamlessly combining — within a single brain — the readout of activity patterns with the readout of structural connectome barcodes [300, 304], transcriptional profiles [301] (e.g., to determine cell type) or other (epi-)genetic signatures [305] which are accessible via high-throughput nucleic acid sequencing.

# 5   Discussion

We have analyzed the physical constraints on scalable neural recording for selected modalities of measurement, data storage, data transmission and power harvesting. Each analysis is based on assumptions – about the brain, device physics, or system architecture – which may be violated. Understanding these assumptions can point towards strategies to work around them, and in some cases we have suggested possible directions for such workarounds. Even valid assumptions about natural brains may be subject to modification through synthetic biology or external perturbation. For example, methods for rapidly removing heat from the brain could work around our assumptions about its natural cooling capacity, supporting a range of highly dissipative recording modalities. Likewise, assumptions about the necessary bandwidth for data transmission could be relaxed if some information is stored locally and read out after the fact.

In some cases, theoretical extensions of our first-order analyses could reveal important insights. The power-bandwidth tradeoffs identified in section 4.4 for electromagnetic data transmission may place limits on the informational throughput of fMRI, for example, or a realistic simulation of heat fluxes in the brain could reveal the true limits of power dissipation. In many other cases, new experiments will be required to move beyond crude estimates of feasibility.

The analysis of physical limits illustrates challenges and opportunities for technology development. While the opportunities can only be touched upon here, and some directions have been treated elsewhere [1, 138, 306], we anticipate further analyses which could explore design spaces in detail. Here we briefly summarize a sampling of new directions suggested by our analysis.

**Electrical recording**   The signal to noise ratio for a voltage sensing electrode imposes limits on the number of neurons per electrode from which signals can be detected and spike-sorted, likely requiring roughly one electrode per 100 neurons. To go beyond this, pure voltage sensing nodes could be augmented with the ability to directionally resolve distinct sources. For example, the 3D motion of a charged nanoparticle in an electric field, or of a dielectric nanoparticle in an electric field gradient, could be monitored at each recording site [307].

**Optical recording**   While light scattering creates severe limitations on optical imaging, embedded optical microscopies could overcome these limits. Embedded optical imaging systems with high signal multiplexing capacity would be desirable, to minimize the required number and size of implanted optical probes.

One option might be to use time-of-flight information to multiplex many sensor readouts into a single optical fiber: this could potentially be realized using time-domain reflectometry techniques, commonly used to determine the positions of defects in optical fibers, coupled to neural activity sensors arranged along the fiber, which would modulate the fiber's local absorption or backscatter [307]. Time-domain reflectometry techniques have already reached 40 μm resolution [308].

Alternatively, novel fluorescent or bio-luminescent activity indicators could in principle relax the limits associated with light scattering, either by enabling efficient two-photon excitation at lower light dosages, or through all-infrared imaging schemes. Infrared bio-luminescence may be a particularly high-value target.

**Delivery**    For both embedded optical and electrical recording strategies, new delivery mechanisms will be needed to scale to whole mammalian brains. Many of the basic parameters for scalable delivery mechanisms are still unknown. For example, can a large number of ultra-thin nano-wire electrodes or optical fibers be delivered via the capillary network? Can cells such as macrophages engulf ultra-miniaturized microchips and transport them into brain tissue? Can the blood brain barrier be locally opened (e.g., using ultrasonic stimulation [309]) to allow targeted delivery of recording probes?

**Intrinsic signals**    The ideal technique would not require exogenous contrast agents or genetically encoded indicators, instead relying on signals intrinsic to neurophysiology. Neurons exhibit few-nano-meter scale [310] membrane displacements (e.g., in response to Maxwell stresses from large local electric field variations) during the action potential [311]. These can be measured using optical interferometry [312], but in principle they could also be monitored acoustically (and related activity-associated membrane swellings have been directly observed by atomic force microscopy [313] in cultured neurons). Sensors could be embedded in or around tissue to transduce the resulting acoustic vibrations into an electrical or optical readout. This could potentially allow recording at larger distances than the ∼130 μm maximum recording radius for a voltage sensing node. Other intrinsic signals include changes in refractive index associated with neural activity, which will modulate the reflection and scattering of light [314]. These intrinsic changes in optical properties can be measured with optical coherence tomography (OCT) [315]. Local metabolic and hemodynamic signatures are also detectable optically, such as hemoglobin oxygenation (e.g., via functional near-infrared spectroscopy [316]) and the partial pressure of oxygen [317, 318]. For minimal invasiveness, diffuse optical tomography uses near-infrared light (600–950 nm), which passes sufficiently readily through the skin and skull to allow imaging of hemodynamics in cortex [319–321], although currently with limited spatial and temporal resolution.

**Data transmission through diffusive media**    Unlike radio-frequency electromagnetics, infrared wavelengths may allow spatially multiplexed data transmissions from embedded recording devices, creating multiple independent channels by taking advantage of the stochasticity of light paths in strongly-scattering tissue. Alternatively, techniques are emerging to dynamically measure and invert the optical scattering matrix of a turbid medium, using pure-optical or hybrid techniques.

**Ultrasound**    Certain wavelengths of ultrasound exhibit potentially-favorable combinations of wavelength (spatial resolution), bandwidth (frequency) and attenuation compared to radio-frequency electromagnetics. Ultrasound could be used as a mechanism for powering and communicating with embedded local recording chips [70]. Novel indicators [322] would likely need to be developed to perform neural activity imaging using pure ultrasound. Hybrid techniques such as photo-acoustic [194] or ultrasound-encoded optical [122] microscopies are also of interest.

**Molecular recording**    For local recording, molecular recording devices could sidestep power constraints on embedded electronics, at the cost of increased engineering complexity. For molecular recording to become practical at temporal resolutions approaching the millisecond scale, sophisticated protein and viral engineering would likely be required to create a high-speed polymerase-based recorder operating in the neuronal cytoplasm. This would also necessitate molecular synchronization or time-stamping mechanisms to maintain phasing between multiple polymerases within a single cell, as well as between different cells.

On the other hand, molecular recording devices operating at slower timescales (e.g., seconds) could perhaps be engineered via more conservative combinations of known mechanisms, such as CREB-mediated signaling to the nucleus [323] or nuclear-localized calcium sensing [324]. In either case, the nucleic acid strands resulting from such molecular recorders could be space-stamped with cell-specific viral connectome barcodes [300] for later readout by bulk sequencing. Alternatively, the ticker tapes could be read within their anatomical contexts by in-situ sequencing, i.e., nucleic acid sequencing performed inside intact tissue [301].

**Combining static and dynamic datasets**    Combining dynamic activity information with static structural or molecular information could allow these datasets to disambiguate one another. For example, a diversity of colors for fluorescent activity indicators (i.e., a form of BrainBow [185] calcium imaging) could ease requirements on spatial separation of optical signals, and the color pattern across cells could be mapped post-mortem at single-cell resolution using in-situ microscopy. Generalizing further, in-situ sequencing enables the extraction of vast quantities of molecular data from fixed tissue, in effect allowing observations with a palette of $4^N$ colors, where $N$ is the length of the nucleic acid polymer. It may be possible to harness this exponential informational resource to enhance the readout of dynamic activity information as well, e.g., through molecular recording.

**MRI** Current MRI is limited by its reliance on intrinsic hemodynamic contrast mechanisms and on rapidly diffusing aqueous protons. Indicators coupling neural activity to spin relaxation rates are being developed to move beyond hemodynamic contrast. Novel excitation and detection schemes that could sensitize MRI to fast, local, intrinsically activity-dependent mechanisms (e.g., cell swelling, neuronal magnetic fields), while filtering out the slower BOLD response, are also of interest and should initially be tested in organisms or slice preparations lacking hemodynamic responses. Detailed computational models of neuronal currents within a tissue voxel (e.g., in the spirit of [71]), and of the resulting mechanical and chemical changes, could be useful for evaluating potential new methods. In principle, MRI could also abandon the use of water protons as the signal sources, although this would pose significant implementation challenges.

**Readout methods** New signal processing frameworks such as compressive sensing could reduce bandwidth requirements and inspire new microscope designs exploiting computational imaging principles [325–328]. Fast readout mechanisms [329] applied to giga-pixel arrays (e.g., the 3.2 giga-pixel CCD camera planned for the Large Synoptic Survey Telescope, which will have ∼1 s readout time) might be adapted to large-scale electrical or optical recording methods. Linear photodiode arrays can achieve 70 kHz line readout rates [330], and many such linear arrays could be read out in parallel. Optoelectronic methods that convert between time, space and frequency representations of signals [331–337] could inspire designs for even faster readouts (e.g., ∼10 MHz frame rates have been demonstrated in brightfield imaging). Although these methods are not directly compatible with fluorescence measurements due to their use of spectral dispersion, related ideas (e.g., beat frequency multiplexing) may enable fluorescence microscopy at rates above that of CCD-based imaging [123, 124], limited ultimately by fluorescence lifetimes, while also exhibiting favorable properties with respect to scattering.

**Alternative modalities** X-ray imaging has been used on live cells [338] and might find use in neural recording if suitable contrast agents could be devised. X-rays interact with electron shells via photoelectric absorption and Compton scattering and with band structure in materials. X-ray phosphors utilize substitutions in an ionic lattice to generate visible or UV light emission upon X-ray absorption [339]. In principle, some of these mechanisms could be engineered as neural activity sensors, e.g., in an absorption-contrast mode suitable for tomographic reconstruction [340]. While tissue damage due to ionizing radiation would ultimately be prohibitive (e.g., on a timescale of minutes [307]), very brief experiments might still be possible.

Likewise, electron spin resonance (ESR) operates at ∼100× higher Larmor frequency compared to proton MRI, which improves polarizability of the spins. Due to Pauli exclusion, use of this technique requires an indicator with unpaired electrons. These can be found in nitrogen vacancy diamond nano-crystals [341] (nano-diamonds), which are also sensitive to voltage [342] and to magnetic fields [343], and are amenable to optical control and fluorescent readout of the spin state (although the 2P cross-section of the $(N-V)^-$ center appears to be relatively low [344]).

**Hybrid systems** New mergers of input, sensing, and readout modalities can work around complex engineering constraints. Electrical or acoustic sensors could be used with optical [345] (e.g., fiber) or ultrasonic readouts and power supplies. An MRI machine could interact with embedded electrical circuits powered by neural activity [346]. Linking electrical recording with embedded optical microscopies or other spatially-resolved methods could circumvent the limits of purely electrical spike sorting. Optical techniques such as holography or 4D light fields could generalize to ultrasound or microwave implementations. Consideration of analogies and synergies between fields suggests a combinatorial space of possibilities.

Our goal here has not been to pick winning technologies (which may not yet have been conceived), but to aid a multi-disciplinary community of researchers in analyzing the problem. The challenge of observing the real-time operation of entire mammalian brains requires a return to first principles, and a fundamental reconsideration of the architectures of neural recording systems. We hope that knowledge of the constraints governing scalable neural recording will enable the invention of entirely new, transformative approaches.

# 6 Acknowledgments

We thank K. Esvelt for helpful discussions on bioluminescent proteins; D. Boysen for help on the fuel cell calculations; R. Tucker and E. Yablonovitch (http://www.e3s-center.org) for helpful discussions on the energy efficiency of CMOS; C. Xu and C. Schaffer for data on optical attenuation lengths; T. Dean and the participants in his CS379C course at Stanford/Google, including Chris Uhlik and Akram Sadek, for helpful discussions and informative content in the discussion notes (http://www.stanford.edu/class/cs379c/); and L. Wood, R. Koene, S. Rezchikov, A. Bansal, J. Lovelock, A. Payne, R. Barish, N. Donoghue, J. Pillow, W. Shih, P. Yin and J. Hewitt for helpful discussions and feedback on earlier drafts.

A. Marblestone is supported by the Fannie and John Hertz Foundation fellowship. D. Dalrymple is supported by the Thiel Foundation. K. Kording is funded in part by the Chicago Biomedical Consortium with support from the Searle Funds at The Chicago Community Trust. E. Boyden is supported by the National Institutes of Health (NIH), the National Science Foundation, the MIT McGovern Institute and Media Lab, the New York Stem Cell Foundation Robertson Investigator Award, the Human Frontiers Science

Program, and the Paul Allen Distinguished Investigator in Neuroscience Award. B. Stranges, B. Zamft, R. Kalhor and G. Church acknowledge support from the Office of Naval Research and the NIH Centers of Excellence in Genomic Science. M. Shapiro is supported by the Miller Research Institute, the Burroughs Wellcome Career Award at the Scientific Interface and the W.M. Keck Foundation.